\documentclass[11pt]{article}
\usepackage{amsmath}
\usepackage{graphicx}
\usepackage{color}
\usepackage[table,xcdraw]{xcolor}
\usepackage{orcidlink}
\usepackage{hyperref}
\hypersetup{colorlinks=true, linkcolor=blue, citecolor=blue, urlcolor=blue}
\usepackage{array}
\usepackage{longtable}
\usepackage{booktabs}
\usepackage{colortbl}
\usepackage{subcaption}
\usepackage{geometry}
\RequirePackage[numbers,sort&compress]{natbib}
\begin{document}
\baselineskip=16pt

\begin{center}
{\Large \textbf{Fingerprints of Loop Quantum Gravity Black Holes with Quintessence Field}}
\end{center}

\vspace{0.1cm}

\begin{center}

{\bf Ahmad Al-Badawi}\orcidlink{0000-0002-3127-3453}\\
Department of Physics, Al-Hussein Bin Talal University, 71111,
Ma'an, Jordan. \\
e-mail: ahmadbadawi@ahu.edu.jo\\

\vspace{0.1cm}

{\bf Faizuddin Ahmed}\orcidlink{0000-0003-2196-9622}\\Department of Physics, Royal Global University, Guwahati, 781035, Assam, India\\
e-mail: faizuddinahmed15@gmail.com\\

\vspace{0.1cm}

{\bf \.{I}zzet Sakall{\i}}\orcidlink{0000-0001-7827-9476}\\
Physics Department, Eastern Mediterranean University, Famagusta 99628, North Cyprus via Mersin 10, Turkey\\
e-mail: izzet.sakalli@emu.edu.tr (Corresponding author)

\end{center}

\vspace{0.1cm}

\begin{abstract}
{\color{black}
In this study, we investigate a static, spherically symmetric black hole (BH) within the framework of Loop Quantum Gravity (LQG) surrounded by quintessence field. Our comprehensive analysis shows that the interplay between quantum corrections and exotic matter produces unique spacetime features, most notably a triple-horizon structure for specific parameter combinations. We derive the metric function incorporating both LQG parameters ($\alpha$, $B$) and quintessence parameters ($c$, $w$), analyzing its implications for horizon structure through embedding diagrams. We examine null and timelike geodesics, calculating photon spheres, effective potentials, and orbital dynamics. Our study demonstrates how quantum and quintessence parameters affect BH shadow size and shape, offering potential observational signatures. Through scalar perturbation analysis, we compute quasinormal modes (QNMs) frequencies, confirming the stability of these hybrid BHs while identifying distinctive spectral characteristics. Finally, using the Gauss-Bonnet (GB) theorem modified approach, we derive an analytical expression for gravitational deflection angles, showing a hierarchical structure of contributions from classical, quintessence, and quantum effects at different distance scales.

} 
\end{abstract}
\section{Introduction} \label{isec1}
{\color{black}
The quest to reconcile quantum mechanics with general relativity (GR) remains one of the most profound challenges in theoretical physics. LQG has emerged as a promising candidate theory in this endeavor, offering a non-perturbative, background-independent approach to quantum gravity \cite{isz01,isz02,isz03}. Unlike string theory, LQG directly quantizes the gravitational field using holonomies and fluxes as fundamental variables, leading to a discrete quantum geometry at the Planck scale \cite{isz04,isz05}. This discrete nature of spacetime, a fundamental prediction of LQG, manifests particularly interestingly in BH physics, where quantum corrections potentially resolve the classical singularity problem that has plagued GR \cite{isz06,isz07}.

In the LQG framework, BHs acquire a modified causal structure with quantum corrections that become significant near the would-be classical singularity. These quantum-corrected BHs often feature multiple horizons-a feature absent in their classical counterparts-with the inner horizon typically shielding a regular core instead of a singularity \cite{isz08,isz09}. The quantization introduces two key parameters into the BH metric: the Barbero-Immirzi parameter $\gamma$, which quantifies the minimum area gap in the theory, and the parameter $\alpha$, which is proportional to $\gamma^3$ times the square of the Planck length $\ell_{pl}^2$ \cite{isz10,isz11}. These parameters, especially in combination, significantly modify the near-horizon physics of BHs, potentially leading to observable consequences in gravitational wave signals and BH shadows \cite{isz12,isz13}. Parallel to these developments in quantum gravity, the discovery of cosmic acceleration has sparked intense interest in exotic matter fields as possible drivers of this phenomenon \cite{isz14,isz15}. The quintessence field (QF), characterized by an equation of state parameter $w_q$ where the pressure $p_q$ relates to the energy density $\rho_q$ via $p_q = w_q\, \rho_q$ with $-1 < w_q < -1/3$, has been particularly successful in modeling dark energy \cite{isz16,isz17}. When incorporated into BH spacetimes, QFs substantially modify their asymptotic structure, often introducing additional horizons at cosmological scales \cite{isz18,isz19}. This field is parametrized by a normalization factor $c$, which quantifies the strength of QF's contribution to the spacetime geometry, alongside the state parameter $w$ determining the equation of state \cite{isz20,isz21}.

BHs in LQG surrounded by QFs represent an intriguing frontier where quantum corrections to near-horizon geometry meet modifications to asymptotic structure induced by exotic matter \cite{isz22,isz23}. Such hybrid spacetimes might harbor unique phenomenological signatures, from modified gravitational lensing to distinctive quasinormal mode (QNM) spectra, potentially providing observational windows into both quantum gravity and dark energy physics simultaneously \cite{isz24,isz25}. In this paper, we introduce a metric describing an LQG BH surrounded by a QF. This metric function encodes both quantum gravity corrections and QF effects, enabling us to study their combined impact on BH physics \cite{isz26,isz27}. Recent observational advancements, particularly the Event Horizon Telescope (EHT)'s imaging of supermassive BHs and LIGO-Virgo-KAGRA's gravitational wave detections, have opened unprecedented opportunities to test modified gravity theories \cite{isz28,isz29}. These observations probe the strong-field regime where both quantum gravity and exotic matter effects might leave detectable imprints \cite{isz30,isz31}. While the parameter space of LQG BHs surrounded by QFs is vast, systematic exploration of their phenomenology could identify distinctive signatures amenable to observational constraints \cite{isz32,isz33}.

In this study, we comprehensively investigate the physics of LQG BHs surrounded by QFs, focusing on their horizon structure, geodesic properties, shadow characteristics, perturbation dynamics, and gravitational lensing behavior. Our analysis reveals how quantum corrections and quintessence effects interplay to produce novel BH phenomenology with potentially observable consequences. We particularly emphasize the triple-horizon structure that emerges for certain parameter combinations, a feature absent in both pure LQG BHs and classical BHs surrounded by quintessence \cite{isz34,isz35}. The motivation for this work is threefold. First, we aim to bridge the gap between quantum gravity phenomenology and exotic matter physics by providing a unified framework for studying their combined effects. Second, we seek to identify distinctive observational signatures that could help constrain both LQG parameters and quintessence properties simultaneously. Finally, we explore the theoretical implications of these hybrid spacetimes for fundamental questions in gravitational physics, including the fate of singularities and the nature of horizons \cite{isz36,isz37}.

The paper is organized as follows. In Section \ref{isec2}, we introduce the LQG-modified BH with QF spacetime and discuss its horizon structure. Section \ref{isec3} examines null and time-like geodesics, analyzing the photon sphere and effective potential. Section \ref{isec4} explores BH shadows and their astrophysical implications. In Sec. \ref{isec5}, we study scalar perturbations and compute the QNMs frequencies. Then, in Sec. \ref{isec6}, we focus on the deflection angle using GBTm. Finally, Sec. \ref{isec7} summarizes our results and outlines future directions.}

\section{LQG BH surrounded by QF} \label{isec2}

The study of the QF as a matter content within GR was carried out by Kiselev \cite{VVK}. He derived a generalisation of the Schwarzschild solution that describes a BH surrounded by a QF, with the associated energy-momentum tensor provided by:
\begin{equation}
    T^{t}_{t}=T^{r}_{r}=\rho_q,\quad T^{\theta}_{\theta}=T^{\phi}_{\phi}=-\frac{1}{2}\,\rho_q\,(3\,w_q+1).\label{pp3}
\end{equation}
where $\rho_q$ denotes the energy density of the QF, and the pressure is related to the density via the equation of state $p_q = w_q \rho_q$, with $w_q$ being the quintessence state parameter. The corresponding line element is given by \cite{VVK}:
\begin{equation}
    ds^2=-\left(1-\frac{2\,M}{r}-\frac{q}{r^{3\,w_q+1}}\right)\,dt^2+\left(1-\frac{2\,M}{r}-\frac{q}{r^{3\,w_q+1}}\right)^{-1}\,dr^2+r^2\,(d\theta^2+\sin^2 \theta\,d\phi^2).\label{pp4}
\end{equation}

\begin{figure}[ht!]
    \centering
    \setlength{\tabcolsep}{0pt} 
    \begin{minipage}{0.32\textwidth}
        \centering
        \includegraphics[width=\textwidth]{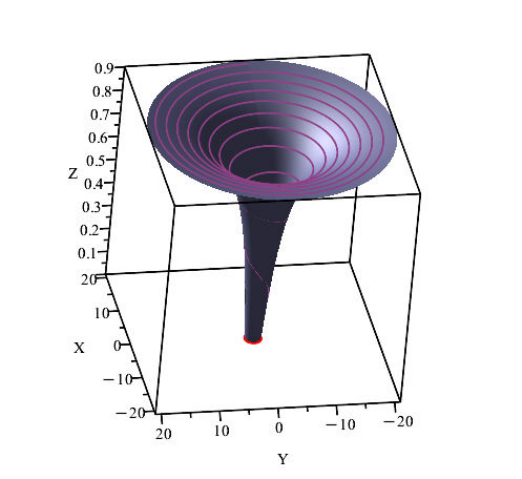}
        \subcaption{[$c=0$, $w=0$, $\alpha=0$] \newline Schwarzschild BH.}
        \label{fig:a}
    \end{minipage}
    \begin{minipage}{0.32\textwidth}
        \centering
        \includegraphics[width=\textwidth]{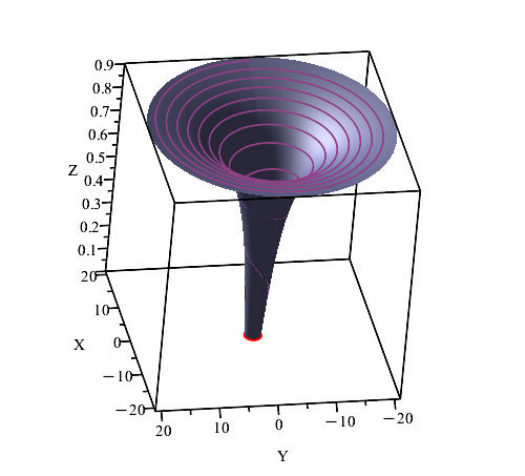}
        \subcaption{[$c=0$, $w=0$, $\alpha=1\times10^{-77}$] \newline LQG BH with $\alpha$ and without QF.}
        \label{fig:b}
    \end{minipage}
    \begin{minipage}{0.32\textwidth}
        \centering
        \includegraphics[width=\textwidth]{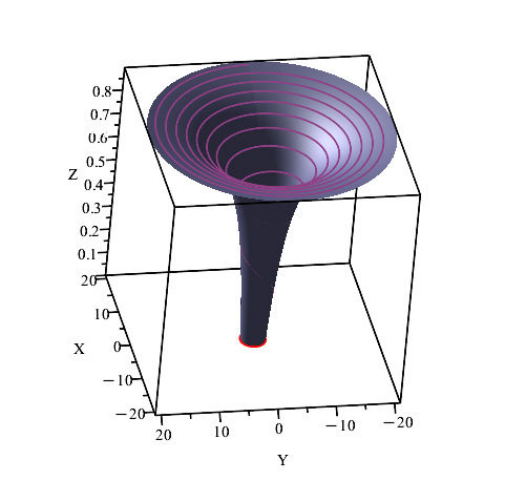}
        \subcaption{[$c=0.06$, $w=0$, $\alpha=0$] \newline LQG BH with $c$ and without QF.}
        \label{fig:c}
    \end{minipage}

    \vspace{0.5em} 

    \begin{minipage}{0.32\textwidth}
        \centering
        \includegraphics[width=\textwidth]{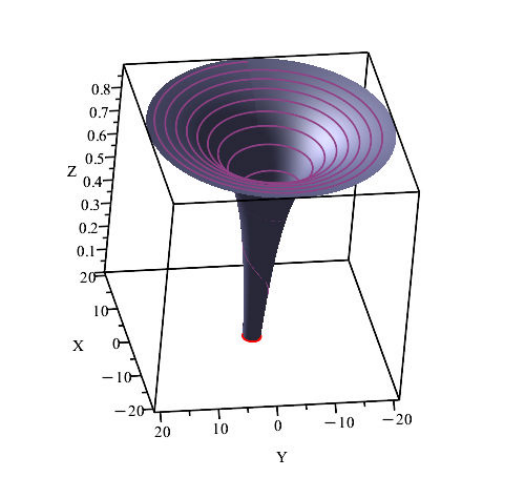}
        \subcaption{[$c=0.06$, $w=0$, $\alpha=1\times10^{-77}$] \newline LQG BH with $\alpha$ and $c$ and without QF.}
        \label{fig:d}
    \end{minipage}
    \begin{minipage}{0.32\textwidth}
        \centering
        \includegraphics[width=\textwidth]{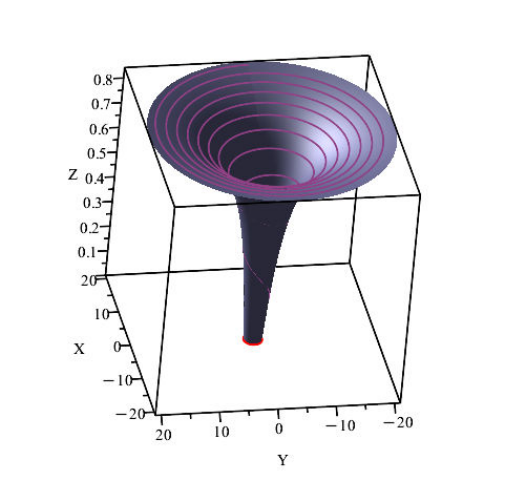}
        \subcaption{[$c=0$, $w=-1/3$, $\alpha=1\times10^{-77}$] \newline LQG BH with $\alpha$ and QF.}
        \label{fig:e}
    \end{minipage}
    \begin{minipage}{0.32\textwidth}
        \centering
        \includegraphics[width=\textwidth]{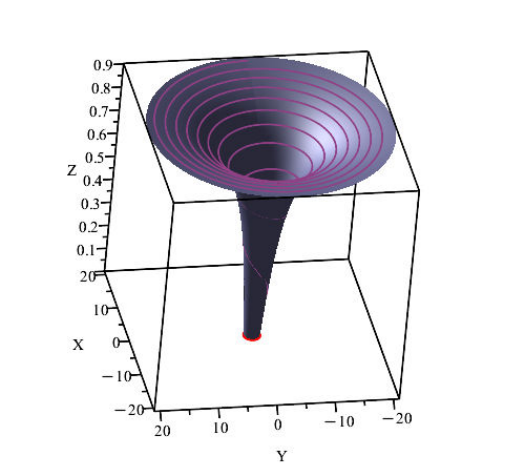}
        \subcaption{[$c=0.06$, $w=-1/3$, $\alpha=1\times10^{-77}$] \newline LQG BH with $\alpha$, $c$, and QF.}
        \label{fig:f}
    \end{minipage}

    \vspace{0.5em}

    \begin{minipage}{0.32\textwidth}
        \centering
        \includegraphics[width=\textwidth]{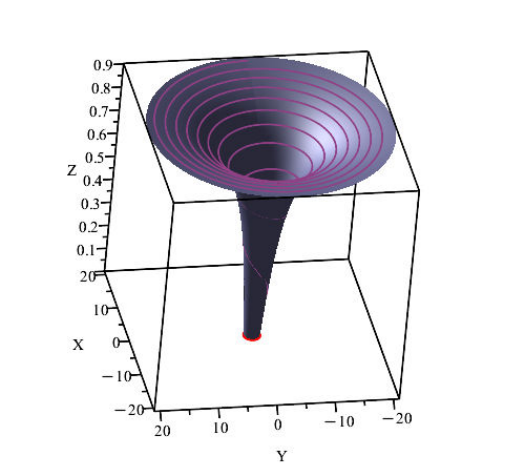}
        \subcaption{[$c=0$, $w=-2/3$, $\alpha=1\times10^{-77}$] \newline LQG BH with $\alpha$ and QF.}
        \label{fig:g}
    \end{minipage}
    \begin{minipage}{0.32\textwidth}
        \centering
        \includegraphics[width=\textwidth]{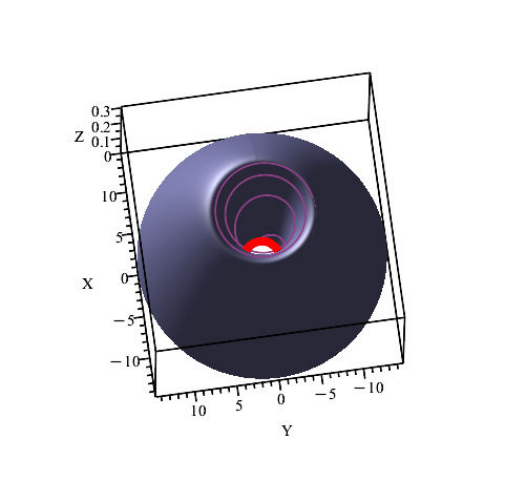}
        \subcaption{[$c=0.06$, $w=-2/3$, $\alpha=0$] \newline LQG BH with $c$ and QF.}
        \label{fig:h}
    \end{minipage}
    \begin{minipage}{0.32\textwidth}
        \centering
        \includegraphics[width=\textwidth]{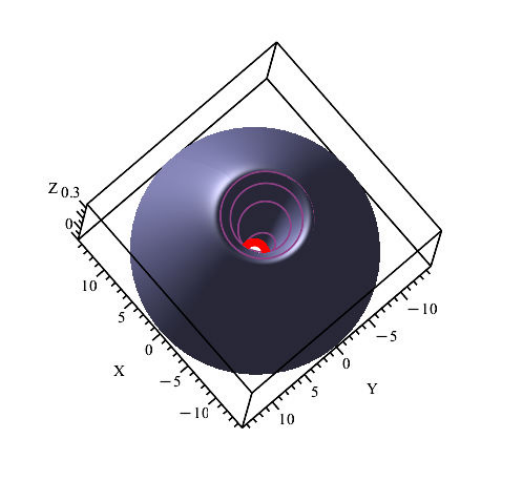}
        \subcaption{[$c=0.06$, $w=-2/3$, $\alpha=1\times10^{-77}$] \newline LQG BH with $\alpha$, $c$, and QF.}
        \label{fig:i}
    \end{minipage}

    \caption{Embedding diagrams of the LQG BH for various parameter values of $w$, $c$, and $\alpha$. The mass and coupling parameters are set to $M=1$ and $B=6\times10^{38}$. The horizons (red rings) are governed by the horizons served in Table \ref{istab}.}
    \label{fig:isfull_embedding}
\end{figure}

On the other hand, the study of spherically symmetric space-times within the framework of LQG is extremely important since it provides insights into the fundamental characteristics of quantum gravity and is intimately related to Birkhoff's theorem \cite{lqg1}. Birkhoff's theorem states that every spherically symmetric vacuum solution to Einstein's equations is necessarily static. 
We extend the LQG BH solutions  by incorporating a QF as the matter source.  Thus, LQG BH surrounded by a QF has the following line element:
\begin{eqnarray}
ds^2=-f(r)\,dt^2+\frac{dr^2}{f(r)}+r^2\,\left(d\theta^2+\sin ^2 \theta \,d \phi^2\right)\label{aa1}
\end{eqnarray}
with the metric function $f(r)$ given by
\begin{eqnarray}
f(r)=1-\frac{2\,M}{r}-\frac{c}{r^{3\,w+1}}+\frac{\alpha}{r^2}\left( \frac{B}{2}+\frac{M}{r}\right)^2,\label{aa2}
\end{eqnarray}
where $B$ is a
 coupling parameter, $\alpha$ parameter is related to the Planck
 length and the Barbero-Immirzi parameter \cite{lqg1,lqg2},  $M$ denotes the  mass
 , $w$ represents the state parameter of the quintessence matter and $c$ serves as a positive normalization factor utilized in the metric equation. Table~\ref{istab} presents the horizon structure of LQG BHs for various combinations of quintessence state parameter $w$, normalization factor $c$, and quantum correction parameter $\alpha$. Several important patterns emerge from these numerical results. When $\alpha=0$ (no quantum corrections), the BH maintains a single horizon regardless of $w$ value, though its radius increases slightly with $c$. However, the introduction of quantum corrections ($\alpha=1\times 10^{-77}$) fundamentally alters this picture, consistently producing a second inner horizon across all parameter combinations. Most notably, the $w=-2/3$ case with both $c=0.06$ and $\alpha=1\times 10^{-77}$ yields a remarkable triple-horizon structure, with horizons at approximately $r=0.66$, $r=1.58$, and $r=14.43$. This extreme case represents the most significant deviation from classical Schwarzschild BH geometry, where the outermost horizon exists at a considerable distance from the central BH. The horizon configurations directly correspond to the red circular rings visible in the embedding diagrams shown in Figure~\ref{fig:isfull_embedding}.
\begin{longtable}{|>{$}c<{$}|>{$}c<{$}|>{$}c<{$}|c|}

\hline
\rowcolor{gray!50}
\textbf{$w$} & \textbf{$c$} & \textbf{$\alpha$} & \text{Horizon(s)} \\
\hline
\endfirsthead
\hline
\rowcolor{gray!50}
\textbf{$w$} & \textbf{$c$} & \textbf{$\alpha$} & \text{Horizon(s)} \\
\hline
\endhead
0 & 0 & 0 & $[2.0]$ \\
\hline
0 & 0 & 1\times 10^{-77} & $[0.6837722340,\ 1.316227766]$ \\
\hline
0 & 0.06 & 0 & $[2.060000000]$ \\
\hline
0 & 0.06 & 1\times 10^{-77} & $[0.6288765776,\ 1.431123422]$ \\
\hline
-\frac{1}{3} & 0 & 0 & $[2.0]$ \\
\hline
-\frac{1}{3} & 0 & 1\times 10^{-77} & $[0.6837722340,\ 1.316227766]$ \\
\hline
-\frac{1}{3} & 0.06 & 0 & $[2.127659574]$ \\
\hline
-\frac{1}{3} & 0.06 & 1\times 10^{-77} & $[0.6463528325,\ 1.481306742]$ \\
\hline
-\frac{2}{3} & 0 & 0 & $[2.0]$ \\
\hline
-\frac{2}{3} & 0 & 1\times 10^{-77} & $[0.6837722340,\ 1.316227766]$ \\
\hline
-\frac{2}{3} & 0.06 & 0 & $[2.324081208,\ 14.34258546]$ \\
\hline
-\frac{2}{3} & 0.06 & 1\times 10^{-77} & $[0.6578301777,\ 1.580363929,\ 14.42847256]$ \\
\hline
\caption{Table of $(w, c, \alpha)$ values of the BH horizons. The other physical parameters are chosen as $M=1$ and $B=6\times 10^{38}$.} \label{istab}
\end{longtable}

\begin{figure}[ht!]
    \centering
    \includegraphics[width=0.65\linewidth]{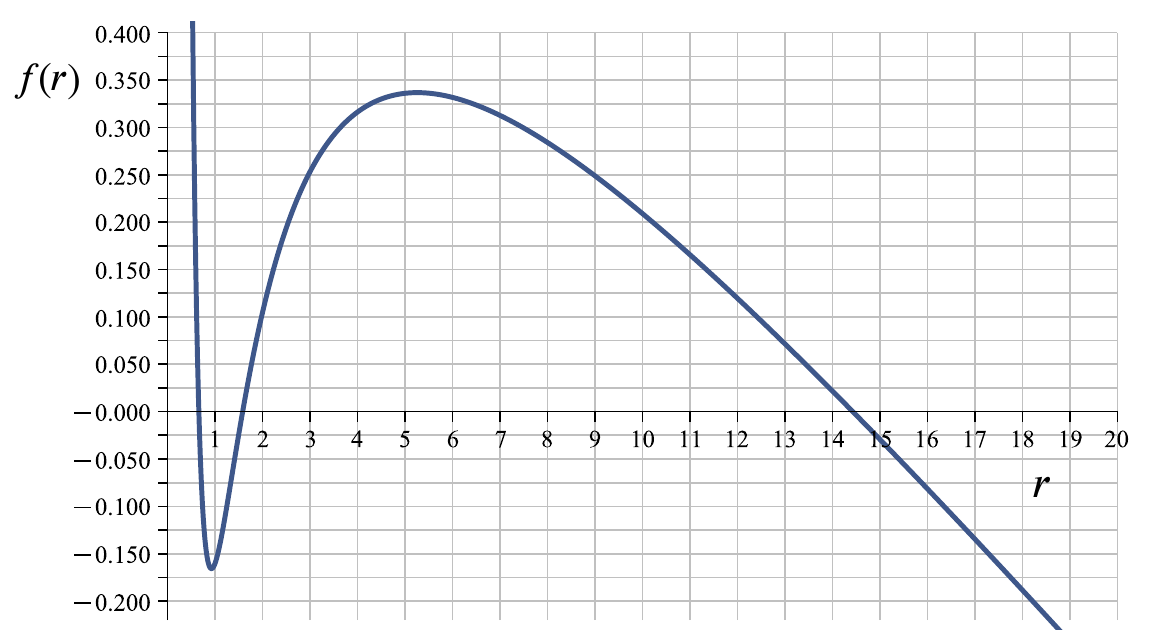}
    \caption{Plot of  $f(r)$ curve where parameters are set as $M = 1$, $c=0.06$, $\alpha=10^{-77}$  and $B=6\times10^{38}$. See also Table \ref{istab}.}
    \label{figa0}
\end{figure}

The embedding diagrams in Fig.~\ref{fig:isfull_embedding} visualize the geometric structure of LQG BHs for various parameter combinations. Panel (a) shows the standard Schwarzschild geometry as a baseline case ($c=0$, $w=0$, $\alpha=0$) with a single horizon at $r=2M$, as confirmed in Table~\ref{istab}. Panels (b)-(i) illustrate how the spacetime curvature is modified by different parameter values. The red rings precisely corresponding to the numerical BH horizon values presented in Table~\ref{istab}. Panel (b) demonstrates the isolated effect of quantum corrections ($\alpha=1\times 10^{-77}$), which splits the single Schwarzschild horizon into two horizons at $r\approx 0.68$ and $r\approx 1.32$. Panel (c) shows how quintessence alone ($c=0.06$) slightly expands the single horizon to $r\approx 2.06$. When both parameters are present in panel (d), their combined effect produces a more pronounced throat structure with horizons at $r\approx 0.63$ and $r\approx 1.43$. The middle row (panels (e)-(g)) explores how different quintessence state parameters ($w$) influence the geometry when quantum corrections are present. For $w=-1/3$ (panel (f)), the horizons appear at $r\approx 0.65$ and $r\approx 1.48$, slightly larger than the $w=0$ case. Notably, as $w$ decreases to $-2/3$ in panel (g), the embedding diagram maintains a similar throat structure with two horizons, consistent with the values in Table~\ref{istab}. The most dramatic effects appear in panels (h) and (i), where $w=-2/3$ and $c=0.06$. As Table~\ref{istab} shows, these combination of parameters generates a third distant (cosmological) horizon (at $r\approx 14.34$ (panel (h)) and $r\approx 14.43$ (panel (i)), resulting in significantly altered embedding geometries. The event (second) horizon, visualized by a red ring far from the central structure, demonstrates that certain parameter combinations can produce spacetime structures fundamentally different from the Schwarzschild BH geometry. Figure~\ref{figa0} illustrates the behavior of the metric function $f(r)$ for an LQG BH surrounded by a quintessence field, with parameters set to $M=1$, $c=0.06$, $\alpha=10^{-77}$, and $B=6\times10^{38}$. The curve reveals critical information about the BH's horizon structure and spacetime properties. The function $f(r)$ crosses the horizontal axis at three distinct points, confirming the triple-horizon structure previously identified in Table~\ref{istab}. 
The innermost horizon occurs at $r \approx 0.66$, followed by an intermediate horizon at $r \approx 1.58$, which together form the typical double-horizon structure of quantum-corrected BHs. However, the most distinctive feature is the third horizon at $r \approx 14.43$, located at a considerable distance from the central BH. Between the second and third horizons, $f(r)$ becomes positive, creating a region where the radial coordinate becomes timelike, similar to the region outside a conventional BH horizon. The function reaches a maximum value of approximately $0.34$ near $r \approx 6$, before gradually declining and eventually crossing the axis again at the third horizon. This behavior demonstrates how the combined effects of LQG corrections and quintessence can substantially modify spacetime structure compared to classical BHs. The region where $f(r) > 0$ between the second and third horizons represents a potentially observable domain with unique physical properties, distinct from both the BH interior and the asymptotic exterior spacetime. This plot provides crucial insight into why the embedding diagrams in Fig.~\ref{fig:isfull_embedding}, particularly panels (h) and (i), exhibit such dramatic differences from the standard Schwarzschild geometry, highlighting the rich phenomenology that emerges from the interplay of quantum gravity effects and exotic matter. 

The above solution (\ref{aa2}) describes the external spacetime of a collapsing ball surrounded by QF, while the FLRW universe describes the spacetime inside the ball 
\begin{equation}
    ds^2_{FLRW}=-dT^2+a^2\left(dR^2+\chi_k ^2(R)  \,d \Omega^2\right),\label{aa3}
\end{equation}
 it was found that the junction conditions lead to \cite{lqg1}
\begin{equation}
 B=-k\chi _{k}^{2}\left( R_{0}\right) =%
\begin{Bmatrix}
-\sin ^{2}\left( R_{0}\right) , & k=1 \\ 
0, & k=0 \\ 
\sinh ^{2}\left( R_{0}\right) , & k=-1%
\end{Bmatrix}   
\end{equation}
where $ R_{0}$ denotes the junction surface. The factor $\alpha$,
 which has the dimension of $r^2$, is given by \cite{lqg1,lqg3}.
\begin{equation}
    \alpha=16\sqrt{3}\,\pi\,\gamma^3\,\ell_{pl} ^2,
\end{equation}
where $\ell_{pl}$ denotes the Planck length, and $\gamma$ is known as the Barbero-Immirzi parameter whose value is set to $\gamma\approx 0.2375$ using BH thermodynamics in LQG \cite{lqg4}.\\ The spacetime represented by the metric function (\ref{aa2}) is defined by two quantum parameters, $\alpha$ and $B$. The factor $\alpha$ which is about $\mathcal{O}$ $(10^{77})$ and dictated by the BH's mass. This leads in limited visible consequences for huge BHs. Similarly, $B$ requires extremely high values (about $10^{38}$ for a solar-mass black hole) to generate discernible effects. In astrophysical black holes, quantum modifications to classical spacetime geometry are minimal, resulting in small observable fingerprints.

\section{Geodesics Analysis of LQG BH surrounded by QF: Null and Time-like Geodesics} \label{isec3}

Geodesic analysis plays a crucial role in BH physics as it describes the motion of massive particles and light in the curved spacetime around BH. By studying geodesics, one can understand key physical phenomena such as the structure of event horizons, the nature of singularities, deflection of light, photon sphere, BH shadows, stability or instability of circular orbits, and the dynamics of accretion disks. Timelike geodesics show how the motion of massive particles influences under gravity, while light-like geodesics help in understanding how light propagates and escapes (or not) from the BH region. This analysis also provides insights into stability, energy extraction mechanisms, and tests of general relativity in strong-field regimes.

In this section, we investigate the null and timelike geodesics around the selected BH solution within the framework of Loop Quantum Gravity in the presence of a quintessence field. We analyze how various parameters-particularly those associated with quantum corrections and the quintessential-affect the geodesic dynamics. To study the geodesic motion of test particles and light, we employ the Lagrangian formalism, a widely adopted approach in the literature (see, for example, Refs. \cite{NPB1, NPB2, CJPHY, EPJC2, PDU1, PDU2, NPB3}, and references therein).

Since the selected spacetime is a static and spherically symmetric, without loss of generality, we consider the geodesic motion in the equatorial plane, defined $\theta=\pi/2$ and $\dot{\theta}=0$. The Lagrangian density function using the metric (\ref{aa1}) is given by
\begin{equation}
    2\,\mathcal{L}=-f(r)\,\dot{t}^2+\frac{\dot{r}^2}{f(r)}+r^2\,\dot{\phi}^2,\label{bb1}
\end{equation}
where dot represents a partial derivative w. r. t. an affine parameter.

From the above Lagrangian function, we observe that the temporal $(t)$ and the azimuthal $(\phi)$ coordinates are the cyclic ones. Therefore, the conserved quantities respectively the energy and angular momentum associated with these cyclic coordinates are given by
\begin{equation}
    \mathrm{E}=f(r)\,\dot{t}\Rightarrow \dot{t}=\frac{\mathrm{E}}{f(r)}.\label{bb2}
\end{equation}
And
\begin{equation}
    \mathrm{L}=r^2\,\dot{\phi} \Rightarrow \dot{\phi}=\frac{\mathrm{L}}{r^2}.\label{bb3}
\end{equation}

With these, geodesic equation for the radial coordinate $r$ using Eq. (\ref{bb1}) can be written as
\begin{equation}
    \dot{r}^2+V_\text{eff}(r)=\mathrm{E}^2\label{bb4}
\end{equation}
which is equivalent to the one-dimensional equation of motion of particles of unit mass having energy $\mathrm{E}^2$ and the effective potential $V_\text{eff}$ given by
\begin{eqnarray}
    V_\text{eff}&=&\left(-\varepsilon+\frac{\mathrm{L}^2}{r^2}\right)\,f(r)=\left(-\varepsilon+\frac{\mathrm{L}^2}{r^2}\right)\,\left(1-\frac{2\,M}{r}-\frac{c}{r^{3\,w+1}}+\frac{\alpha}{r^2}\left( \frac{B}{2}+\frac{M}{r}\right)^2\right). \label{bb5}
\end{eqnarray}
where $f(r)$ is given in (\ref{aa2}), $\varepsilon=0$ for null geodesics and $-1$ for time-like particles.

The expression (\ref{bb5}) shows that the effective potential for both light-like and time-like particles is affected by several parameters. These include quantum corrections represented by the parameter $\xi$, the cosmic string parameter $\alpha$, the normalization constant $c$ of the QF, and the state parameter $w$. Moreover, the potential for light and time-like particles is modified by angular momentum $\mathrm{L}$. Together, these parameters modify the gravitational field of the chosen BH solution, which in turn leads to modifications in the motion of test particles. 

Below, we discuss in detail the motions of light-like and time-like particles and analyze the outcomes.

\subsection{Null Geodesics: photon sphere, force, and trajectory}

Null geodesics describe the trajectories of massless particles, such as photons, in curved spacetime. These paths are governed by the BH's gravitational field and can be effectively analyzed using an effective potential derived from the spacetime metric. This effective potential framework allows for the investigation of various geometric and physical properties, including the deflection of light, the structure and size of the BH shadow, dynamics of photon particles, photon sphere radius, and the stability of circular photon orbits.

For light-like particles, $\varepsilon=0$ and hence, the effective potential from Eq. (\ref{bb5}) reduces as,
\begin{equation}
    V_\text{eff}=\frac{\mathrm{L}^2}{r^2}\,\left(1-\frac{2\,M}{r}-\frac{c}{r^{3\,w+1}}+\frac{\alpha}{r^2}\left( \frac{B}{2}+\frac{M}{r}\right)^2\right).\label{cc1}
\end{equation}

\begin{figure}[ht!]
    \centering
    \subfloat[$c=0.01, B=0.5$]{\centering{}\includegraphics[width=0.45\linewidth]{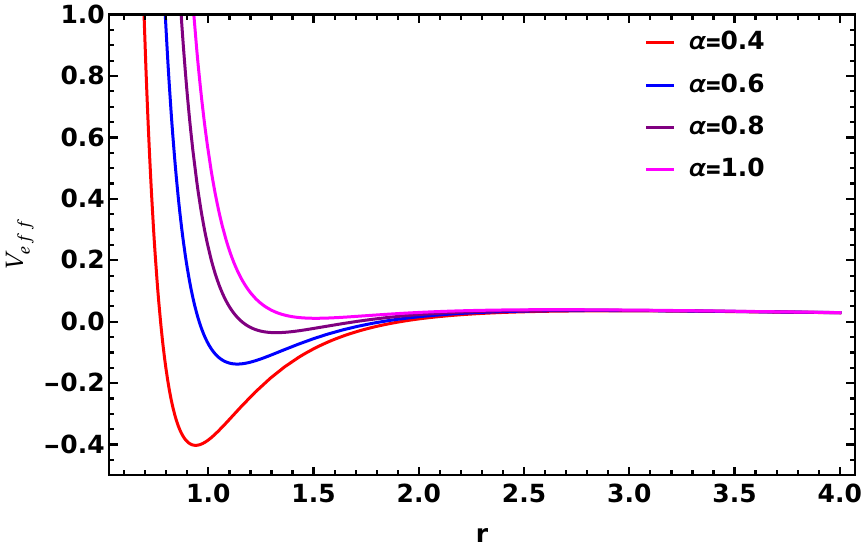}}\quad\quad
    \subfloat[$c=0.01, \alpha=0.5$]{\centering{}\includegraphics[width=0.45\linewidth]{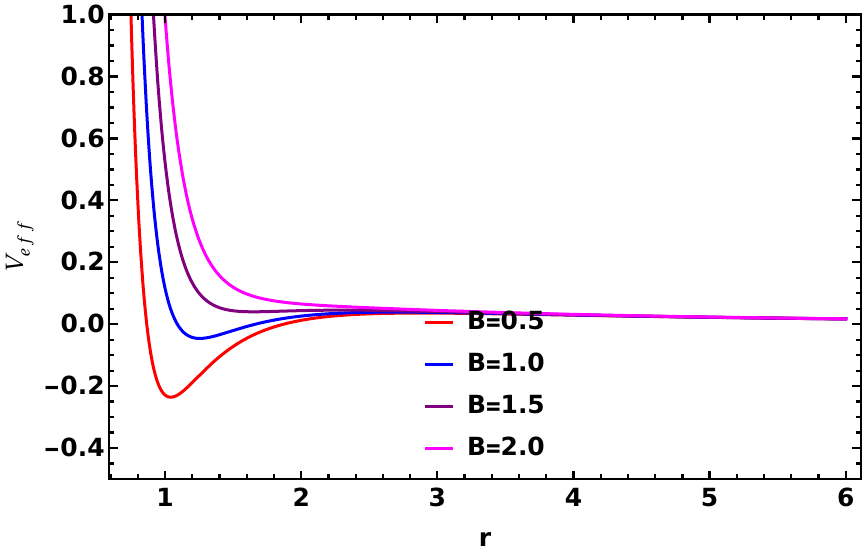}}\\
    \subfloat[$\alpha=0.5=B$]{\centering{}\includegraphics[width=0.45\linewidth]{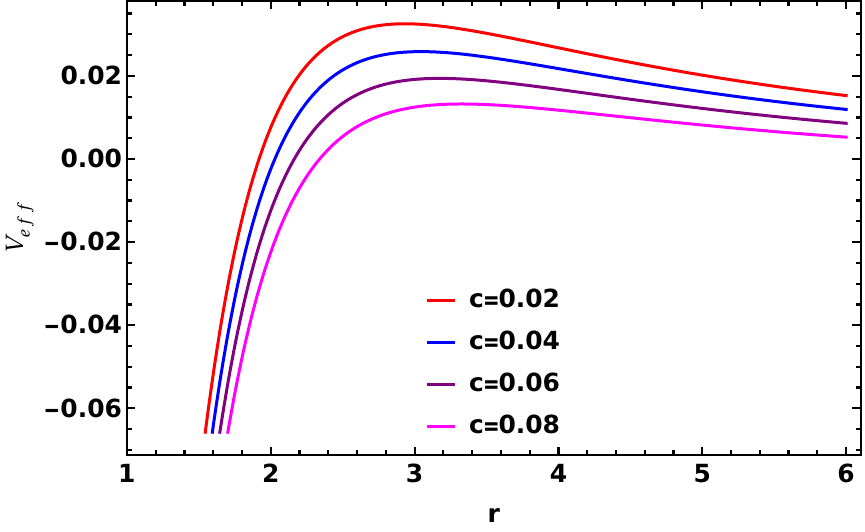}}\quad\quad
    \subfloat[]{\centering{}\includegraphics[width=0.45\linewidth]{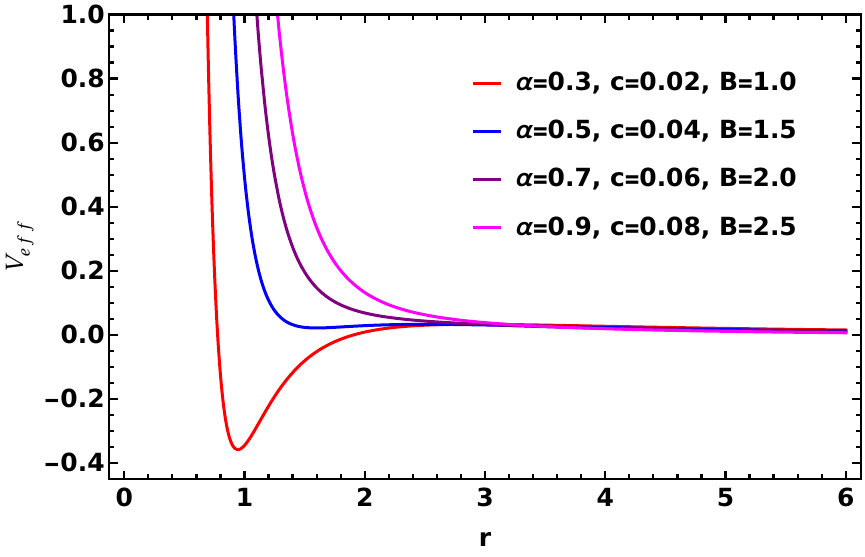}}
    \caption{The behavior of the effective potential for null geodesics as a function of $r$ for different values of $\alpha, B$ and $c$. here, we set $M=1, L=1, w=-2/3$.}
    \label{fig:null-potential}
\end{figure}

In Figure \ref{fig:null-potential}, we present a series of plots illustrating the behavior of the effective potential for null geodesics as the parameters $\alpha$, $B$, $c$, and their combinations vary. In all panels except panel (c), we observe that the effective potential increases with increasing values of these parameters or their combinations. An increase in the effective potential with rising parameters (individually or combination) value implies a stronger gravitational influence that modifies the paths of light rays-typically making it harder for them to escape or penetrate into the BH region. This affects observable phenomena such as light deflection, photon spheres, and shadow profiles. In contrast, panel (c) shows that the effective potential decreases as the value of the quintessential parameter $c$ increases keeping fixed other parameters.

To study circular null geodesics of radius $r=r_c$, we have the conditions $\dot{r}=0$ and $\ddot{r}=0$. Thereby, using Eq. (\ref{bb4}), we find the following two important relations
\begin{equation}
    \mathrm{E}^2=V_\text{eff}(r)=\frac{\mathrm{L}^2}{r^2}\,\left(1-\frac{2\,M}{r}-\frac{c}{r^{3\,w+1}}+\frac{\alpha}{r^2}\left( \frac{B}{2}+\frac{M}{r}\right)^2\right).\label{cc2}
\end{equation}
And
\begin{equation}
    V_\text{eff}'(r)=0.\label{cc3}
\end{equation}

The first relation gives us the critical impact parameter at radius $r=r_c$ for photon particles and is given by
\begin{equation}
    \beta_c=\frac{r_c}{\sqrt{1-\frac{2\,M}{r_c}-\frac{c}{r^{3\,w+1}_c}+\frac{\alpha}{r^2_c}\left( \frac{B}{2}+\frac{M}{r_c}\right)^2}}.\label{cc4}
\end{equation}

From expression given in Eq. (\ref{cc4}), it becomes evident that the critical impact parameter $\beta_c$ is influenced by several factors present in the BH space-time. These include the coupling parameter $B$, the parameter $\alpha$, QF parameters $(c, w)$. Additionally, the BH mass $M$ also effect this impact parameter. Depending on the relationship between the critical impact parameter $\beta_c$ and impact parameter $\beta=\mathrm{L}/\mathrm{E}$, the photon particles may come closer or move away from the BH region and/or orbiting in a circular paths. For a specific state parameter $w=-2/3$, we find the critical impact parameter
\begin{equation}
    \beta_c=\frac{r_c}{\sqrt{1-\frac{2\,M}{r_c}-c\,r_c+\frac{\alpha}{r^2_c}\left( \frac{B}{2}+\frac{M}{r_c}\right)^2}}.\label{cc4aa}
\end{equation}

The second relation given in Eq. (\ref{cc3}) determines the photon sphere radius $r=r_\text{ph}$ satisfying the following equation
\begin{equation}
    1-\frac{3\,M}{r}-\frac{c\,(3\,w+3)}{2\,r^{3\,w+1}}+\frac{\alpha}{2\,r^2}\,\left(B+\frac{3\,M}{r}\right)\,\left(B+\frac{2\,M}{r}\right)=0.\label{cc5}
\end{equation}

As for example, setting the state parameter $w=-2/3$, from Eq. (\ref{cc5}) we find
\begin{equation}
    1-\frac{3\,M}{r}-\frac{c\,r}{2}+\frac{\alpha}{2\,r^2}\,\left(B+\frac{3\,M}{r}\right)\,\left(B+\frac{2\,M}{r}\right)=0.\label{cc6}
\end{equation}

The above equation can be expressed as $r^5+c_4\,r^4+c_3\,r^3+c_2\,r^2+c_1\,r+c_0=0$, where $c_i,\,i=0\cdots 4$ are various order coefficients. The exact expression for photon sphere radius $r=r_\text{ph}$ is complicated. However, one can numerically determine this radius by choosing values of $c$, $B$, $\alpha$ and $M$. From expression (\ref{cc6}), it is clear that several factors, such as the coupling parameter $B$, the parameter $\alpha$, QF normalization constant $c$. Additionally, the BH mass $M$ also alter this radius.

\begin{figure}[ht!]
    \centering
    \subfloat[$c=0.01, B=0.5$]{\centering{}\includegraphics[width=0.45\linewidth]{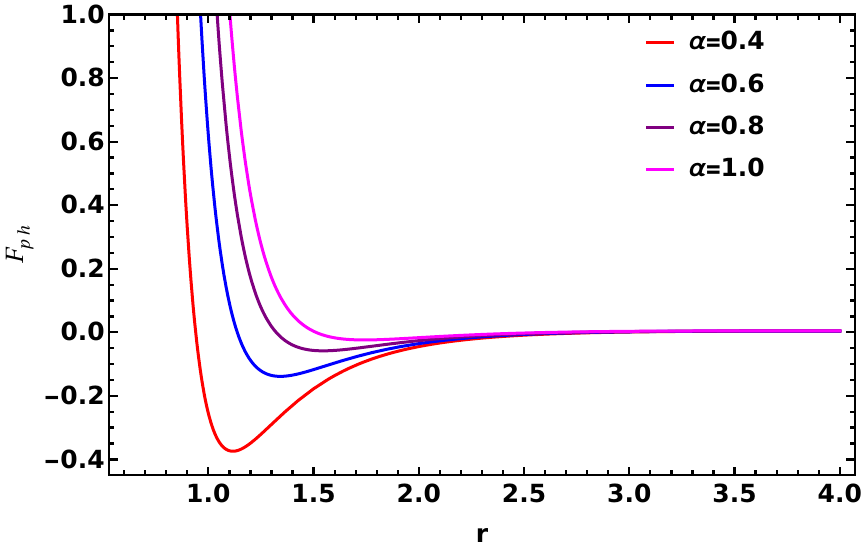}}\quad\quad
    \subfloat[$c=0.01, \alpha=0.5$]{\centering{}\includegraphics[width=0.45\linewidth]{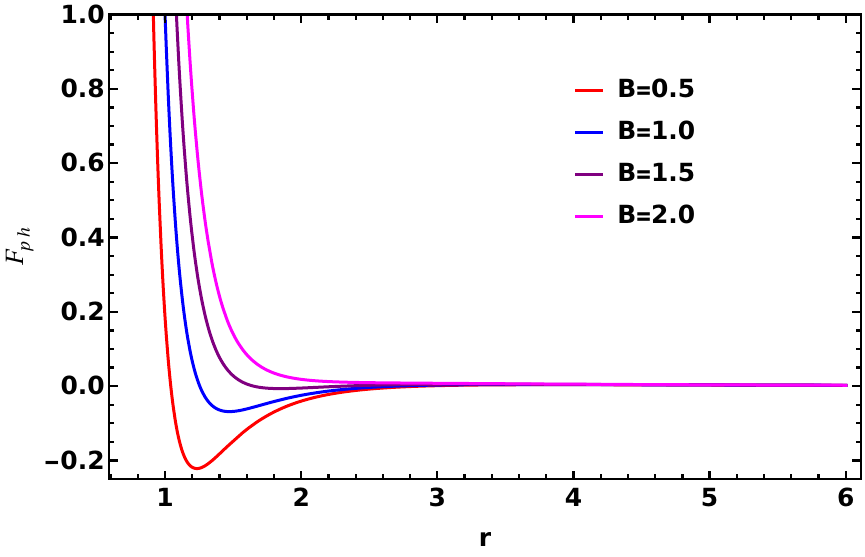}}\\
    \subfloat[$\alpha=1.5, B=0.5$]{\centering{}\includegraphics[width=0.46\linewidth]{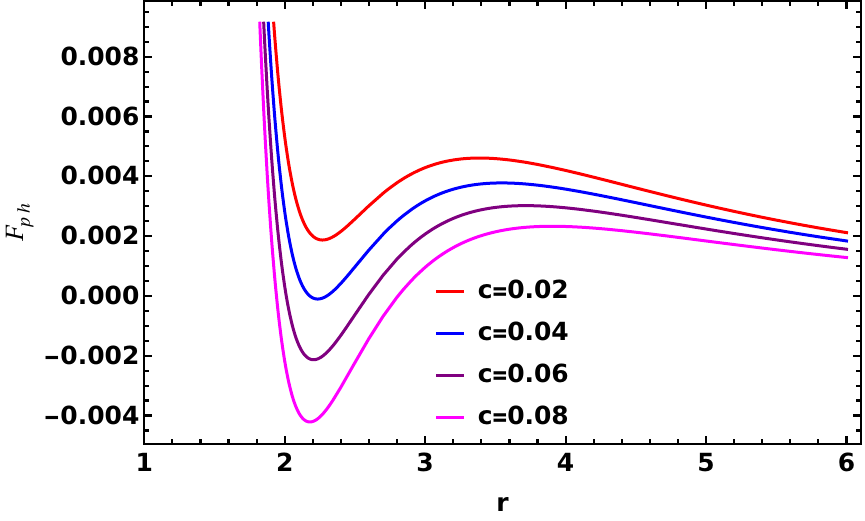}}\quad\quad
    \subfloat[]{\centering{}\includegraphics[width=0.45\linewidth]{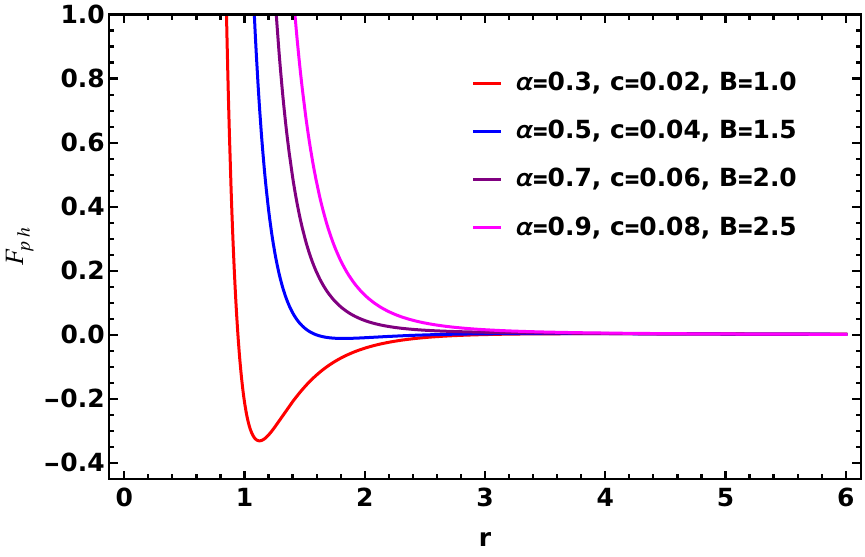}}
    \caption{The behavior of the force acting on photon particles as a function of $r$ for different values of $\alpha, B$ and $c$. here, we set $M=1, L=1, w=-2/3$.}
    \label{fig:force}
\end{figure}

We now calculate the force acting on photon particles under the influence of the gravitational field generated by the BH. Furthermore, we examine how different factors affect the dynamics of photons as they interact with the gravitational field of the BH. The force on photons is related with effective potential $V_\text{eff}(r)$ for null geodesics as
\begin{equation}
    \mathrm{F}_\text{ph}=-\frac{1}{2}\,\frac{\partial V_\text{eff}}{\partial r}.\label{cc7}
\end{equation}

Substituting the effective potential from Eq. (\ref{cc1}) results
\begin{equation}
    \mathrm{F}_\text{ph}=\frac{\mathrm{L}^2}{r^3}\,\left[1-\frac{3\,M}{r}-\frac{c\,(3\,w+3)}{2\,r^{3\,w+1}}+\frac{\alpha}{2\,r^2}\,\left(B+\frac{3\,M}{r}\right)\,\left(B+\frac{2\,M}{r}\right)\right].\label{cc8}
\end{equation}

From expression given in Eq. (\ref{cc8}), it becomes evident that the force $\mathrm{F}_\text{ph}$ acting on photon particles is influenced by several factors present in the BH space-time. These include the coupling parameter $B$, the parameter $\alpha$, QF parameters $(c, w)$. Additionally, the angular momentum $\mathrm{L}$ and BH mass $M$ also influence the dynamics of photon particles.

As for example, setting the state parameter $w=-2/3$ into the Eq. (\ref{cc8}), we find the force on photon particles given by
\begin{equation}
    \mathrm{F}_\text{ph}=\frac{\mathrm{L}^2}{r^3}\,\left[1-\frac{3\,M}{r}-c\,r+\frac{\alpha}{r^2}\,\left(B+\frac{3\,M}{r}\right)\,\left(B+\frac{2\,M}{r}\right)\right].\label{cc9}
\end{equation}

In Figure \ref{fig:force}, we present a series of plots illustrating the behavior of the force acting on photon particles in the gravitational field as the parameters $\alpha$, $B$, $c$, and their combinations vary. In all panels except panel (c), we observe that the force increases with increasing values of these parameters or their combinations. An increase in the force with rising parameter values-either individually or in combination-indicates a stronger gravitational influence, which may manifest as either enhanced attraction or repulsion depending on the nature of the parameter. Specifically, a transition in the force from negative to positive values suggests that photon particles experience a net outward push, implying a tendency to move away from the BH region, potentially due to repulsive effects introduced by the modified gravity or matter fields. In contrast, panel (c) shows that the force decreases as the value of the quintessential parameter $c$ increases, with other parameters held fixed. This behavior suggests that higher values of $c$ enhance the overall attractive nature of the gravitational field, leading to a stronger confinement of photon particles near the BH. In this scenario, the quintessence field acts to deepen the gravitational potential well, making it more difficult for photons to escape and effectively binding them more tightly to the black hole's vicinity.

Now, we focus into the photon trajectories and show how different factors affect the photon trajectory in the vicinity of the BH. For null geodesics, using Eqs. (\ref{bb3}) and (\ref{bb4}) we define the following 
\begin{equation}
    \frac{\dot{r}^2}{\dot{\phi}^2}=\left(\frac{dr}{d\phi}\right)^2=r^4\,\left[\frac{1}{\beta^2}-\frac{1}{r^2}\,\left\{1-\frac{2\,M}{r}-\frac{c}{r^{3\,w+1}}+\frac{\alpha}{4\,r^2}\left(B+\frac{2\,M}{r}\right)^2\right\}\right],\label{cc10}
\end{equation}
where $\beta=\mathrm{L}/\mathrm{E}$ is the impact parameter for photon particles.

In order to simplify the above equation, setting the state parameter $w=-2/3$ and a change of variable via $u=\frac{1}{r}$ into the eq. (\ref{cc10}) results
\begin{equation}
    \left(\frac{du}{d\phi}\right)^2+u^2=\frac{1}{\beta^2}+2\,M\,u^3+c\,u-\frac{\alpha\,u^4}{4}(B+2\,M\,u)^2.\label{cc11}
\end{equation}
Differentiating w. r. t. $\phi$ and after simplification, we find the following trajectory equation
\begin{equation}
    \frac{d^2u}{d\phi^2}+u=\frac{c}{2}+3\,M\,u^2-\frac{\alpha\,u^3}{2}\,(B+2\,M\,u)\,(B+3\,M\,u).\label{cc13}
\end{equation}
Equation (\ref{cc13}) is the second-order non-linear differential equation for photon paths traversing around the BH.

Now, we discuss further stability of circular orbits of radius $r=r_c$. For that, we need to determine a physical quantity called the Lyapunov exponent for circular null geodesics given by \cite{VC}
\begin{equation}
    \lambda^\text{null}_{L}=\sqrt{-\frac{V''_\text{eff}(r)}{2\,\dot{t}^2}}\Big{|}_{r=r_c},\label{cc14}
\end{equation}
where $\dot{t}=\frac{\mathrm{E}}{f(r)}$ and for null geodesics we have $\frac{\mathrm{E}}{\mathrm{L}}=\frac{f(r_c)}{r^2_c}$.

\begin{figure}[ht!]
    \centering
    \subfloat[$c=0.01,B=0.1$]{\centering{}\includegraphics[width=0.45\linewidth]{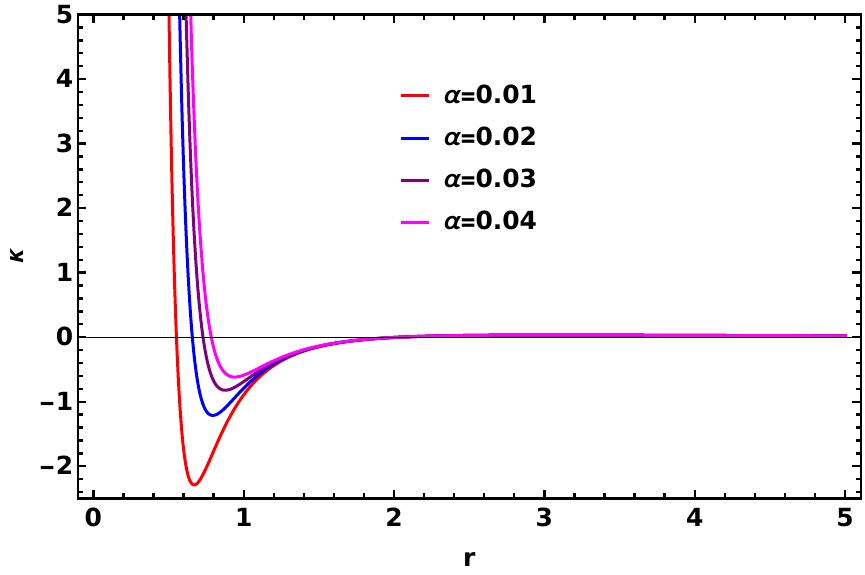}}\quad\quad
    \subfloat[$\alpha=0.1=\beta$]{\centering{}\includegraphics[width=0.46\linewidth]{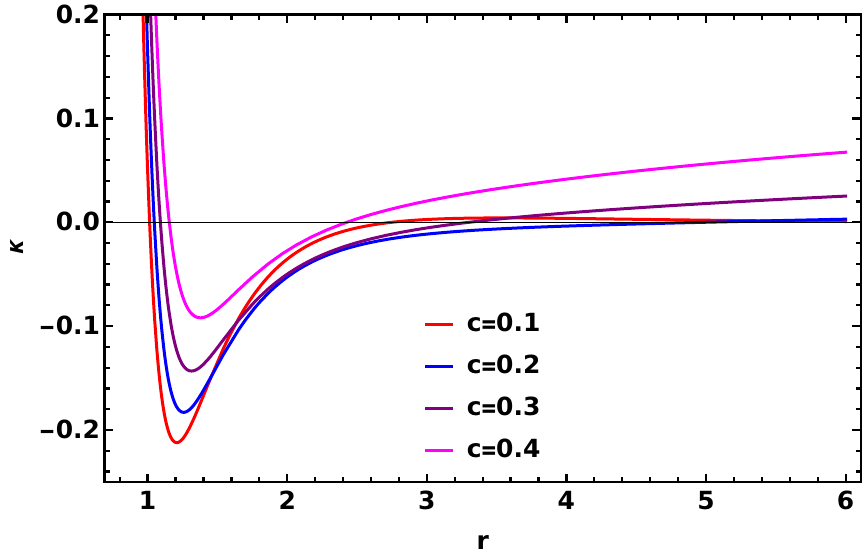}}\\
    \subfloat[$c=0.01$]{\centering{}\includegraphics[width=0.45\linewidth]{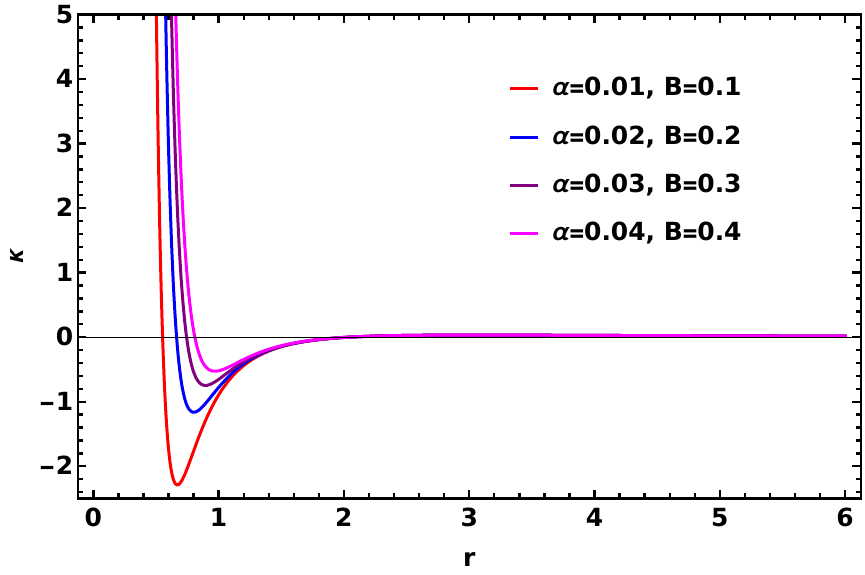}}\quad\quad
    \subfloat[$\beta=0.1$]{\centering{}\includegraphics[width=0.45\linewidth]{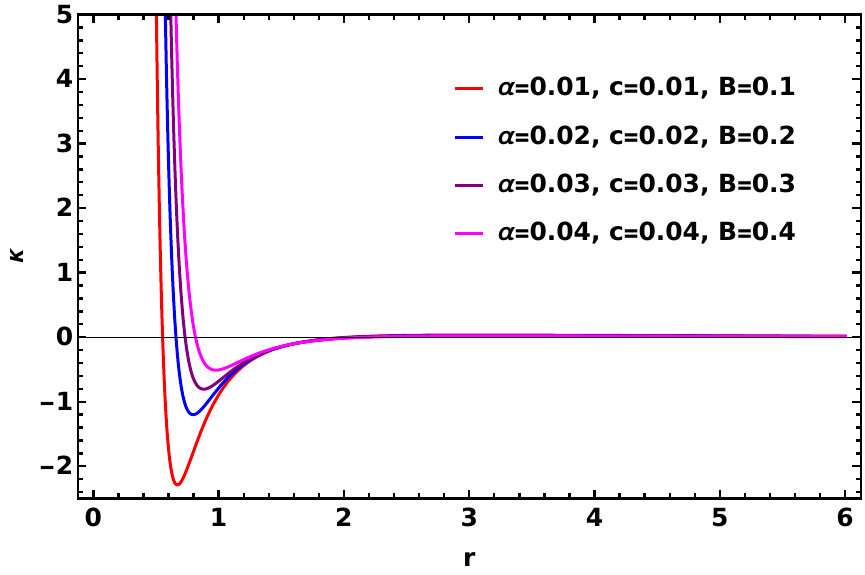}}
    \caption{The behavior of the Lyapunov exponent $\kappa=\left(\lambda^\text{null}_{L}\right)^2$ for circular null orbits as a function of $r_c$ for different values of $\alpha, B$ and $c$. Here, we set $M=1, w=-2/3$.}
    \label{fig:lyapunov}
\end{figure}

Substituting the effective potential from Eq. (\ref{cc1}) and after simplification, we find
\begin{equation}
    \lambda^\text{null}_{L}=\sqrt{1-\frac{2\,M}{r}-\frac{c}{r^{3\,w+1}}+\frac{\alpha}{r^2}\left(\frac{B}{2}+\frac{M}{r}\right)^2}\,\sqrt{\frac{1}{r^2}+\frac{c\,((3\,w+1)\,(3\,w+2)-2)}{2\,r^{3\,w+3}}-\frac{\alpha}{r^4}\,\left(\frac{B^2}{2}+\frac{9\,M^2}{r^2}+\frac{5\,M\,B}{r}\right)}\Big{|}_{r=r_c}.\label{cc15}
\end{equation}

From the expression given in Eq. (\ref{cc15}), it becomes evident that the Lyapunov exponent for null geodesics is influenced by several factors present in BH space-time geometry. These include the coupling parameter $B$, the parameter $\alpha$, QF parameters $(c, w)$. Additionally, the BH mass $M$ also influence this physical quantity.

As for example, setting the state parameter $w=-2/3$ into the Eq. (\ref{cc15}), we find the Lyapunov exponent 
\begin{equation}
    \lambda^\text{null}_{L}=\sqrt{1-\frac{2\,M}{r_c}-c\,r_c+\frac{\alpha}{r^2_c}\left(\frac{B}{2}+\frac{M}{r}\right)^2}\,\sqrt{\frac{1}{r^2_c}-\frac{c}{r_c}-\frac{\alpha}{r^4_c}\,\left(\frac{B^2}{2}+\frac{9\,M^2}{r^2_c}+\frac{5\,M\,B}{r_c}\right)}.\label{cc16}
\end{equation}

In Figure \ref{fig:lyapunov}, we present a series of plots illustrating the behavior of the squared Lyapunov exponent for circular null orbits as the parameters $\alpha$, $B$, $c$, and their combinations vary. In all panels, we observe that the squared Lyapunov exponent remains negative within a specific range of the radial coordinate, $r = r_c$, as these parameters or their combinations increase. A negative value of the squared Lyapunov exponent corresponds to an \textit{imaginary Lyapunov exponent}, which indicates that the circular null orbits are \textit{radially stable}. In such cases, small perturbations do not cause nearby photon trajectories to diverge exponentially; instead, they undergo bounded oscillations around the circular path. This behavior implies the existence of \textit{stable photon orbits}, a feature that is generally absent in classical BH solutions such as Schwarzschild or Kerr. The presence of such stable null orbits may point to \textit{non-standard or exotic spacetime geometries}, potentially arising from modifications to general relativity or the inclusion of additional fields like quintessence.

\begin{figure}[ht!]
    \centering
    \subfloat[$c=0.01, B=0.5$]{\centering{}\includegraphics[width=0.45\linewidth]{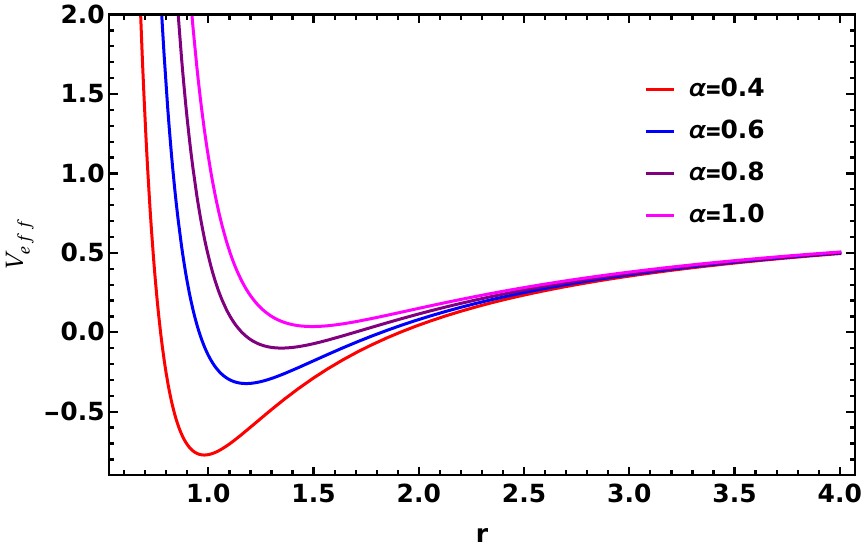}}\quad\quad
    \subfloat[$c=0.01, \alpha=0.5$]{\centering{}\includegraphics[width=0.45\linewidth]{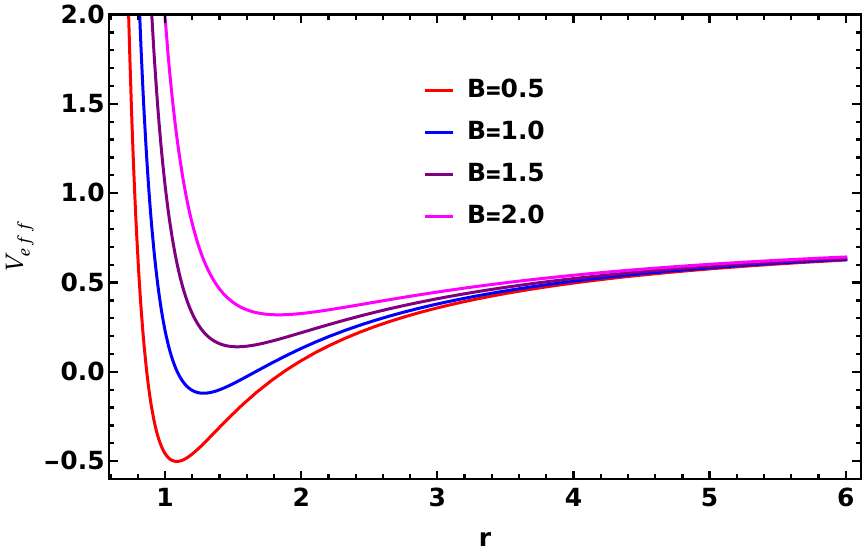}}\\
    \subfloat[$\alpha=0.5=B$]{\centering{}\includegraphics[width=0.45\linewidth]{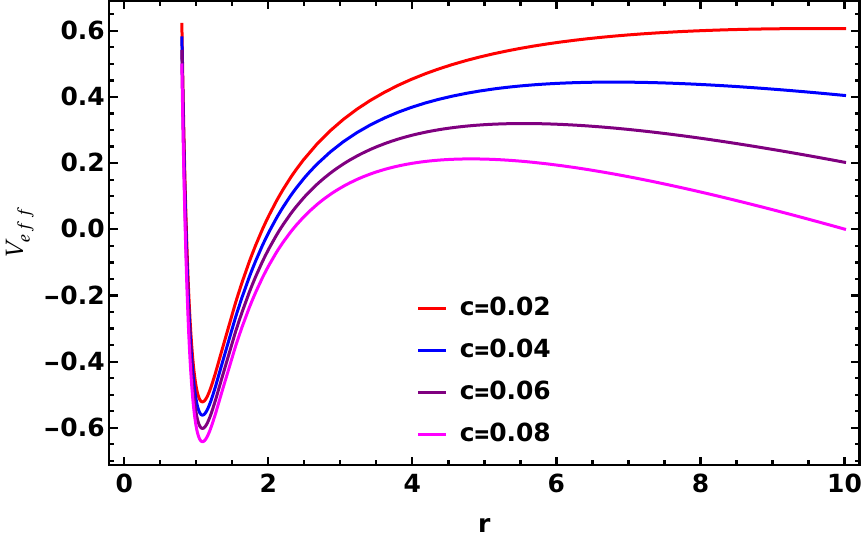}}\quad\quad
    \subfloat[]{\centering{}\includegraphics[width=0.45\linewidth]{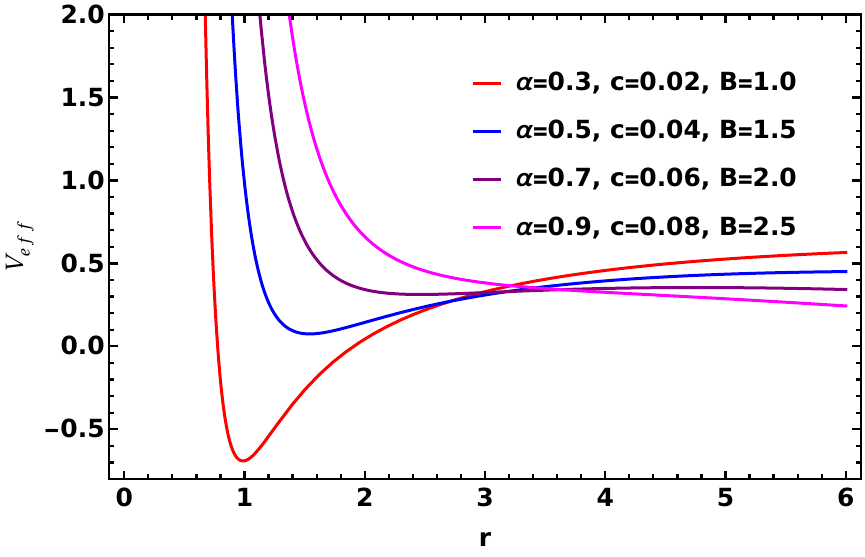}}
    \caption{The behavior of the effective potential for time-like geodesics as a function of $r$ for different values of $\alpha, B$ and $c$. here, we set $M=1, L=1, w=-2/3$.}
    \label{fig:timelike-potential}
\end{figure}

Finally, the coordinate angular velocity is defined by \cite{VC}
\begin{equation}
    \Omega^\text{null}=\frac{\dot{\phi}}{\dot{t}}\Big{|}_{r=r_c}=\frac{f(r_c)}{r^2_c}\,\frac{\mathrm{L}}{\mathrm{E}}=\frac{\sqrt{f(r_c)}}{r_c}=\frac{\sqrt{1-\frac{2\,M}{r_c}-\frac{c}{r^{3\,w+1}_c}+\frac{\alpha}{r^2_c}\left( \frac{B}{2}+\frac{M}{r_c}\right)^2}}{r_c},\label{cc17}
\end{equation}
where we have used the relation (\ref{cc2}).

From the expression given in Eq. (\ref{cc17}), it becomes evident that the coordinate angular velocity in circular orbits is influenced by several factors: the coupling parameter $B$, the parameter $\alpha$, QF parameters $(c, w)$. Additionally, the BH mass $M$ also influence this coordinate velocity.

\subsection{Time-like geodesics}

Time-like geodesics describe the trajectories of massive particles (e.g., stars, planets, or observers) moving in the curved spacetime around a BH. Depending on their specific energy and angular momentum, these particles can fall into the BH, escape to infinity, or remain in bound or circular orbits. Among these, circular orbits are particularly significant for understanding physical phenomena such as accretion disks and the innermost stable circular orbit (ISCO). The effective potential for time-like geodesics serves as a tool to analyze the motion of particles in the vicinity of the BH. It allows us to identify stable and unstable orbits, as well as regions of allowed motion. Furthermore, by analyzing the effective potential at large radial distances ($r \gg r_\text{h}$, where $r_\text{h}$ is the event horizon radius), one can determine the orbital velocity of massive particles in the asymptotically flat region of the BH space-time.
  
For massive test particles, $\varepsilon=-1$, the effective potential from (\ref{bb5}) reduces as,
\begin{equation}
    V_\text{eff}=\left(1+\frac{\mathrm{L}^2}{r^2}\right)\,\left(1-\frac{2\,M}{r}-\frac{c}{r^{3\,w+1}}+\frac{\alpha}{r^2}\left( \frac{B}{2}+\frac{M}{r}\right)^2 \right).\label{dd1}
\end{equation}
And the geodesics equation for the radial coordinate $r$ will be
\begin{equation}
    \dot{r}=\sqrt{\mathrm{E}^2-V_\text{eff}},\label{dd2}
\end{equation}
where $V_\text{eff}$ is given in Eq. (\ref{dd1}). 

In Figure \ref{fig:timelike-potential}, we present a series of plots illustrating the behavior of the effective potential for time-like geodesics as the parameters $\alpha$, $B$, $c$, and their combinations vary. In all panels except panel (c), we observe that the effective potential increases with increasing values of these parameters or their combinations. An increase in the effective potential with rising parameters (individually or combination) value implies a stronger gravitational influence that modifies the paths of massive particles-typically making it harder for them to escape or penetrate into the BH region. In contrast, panel (c) shows that the effective potential decreases as the value of the quintessential parameter $c$ increases keeping fixed other parameters.

\begin{figure}
    \centering
    \subfloat[$c=0.01, B=0.1$]{\centering{}\includegraphics[width=0.4\linewidth]{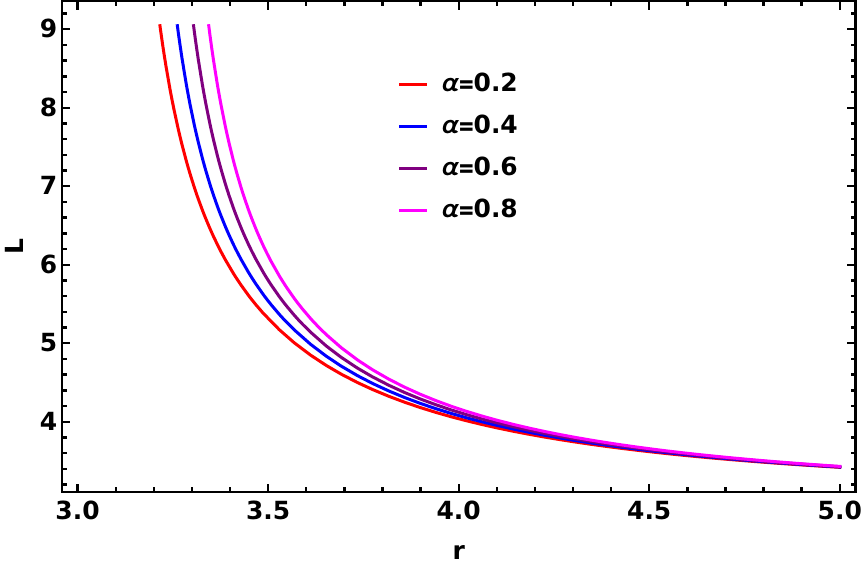}}\quad\quad\quad
    \subfloat[$c=0.01, B=0.1$]{\centering{}\includegraphics[width=0.41\linewidth]{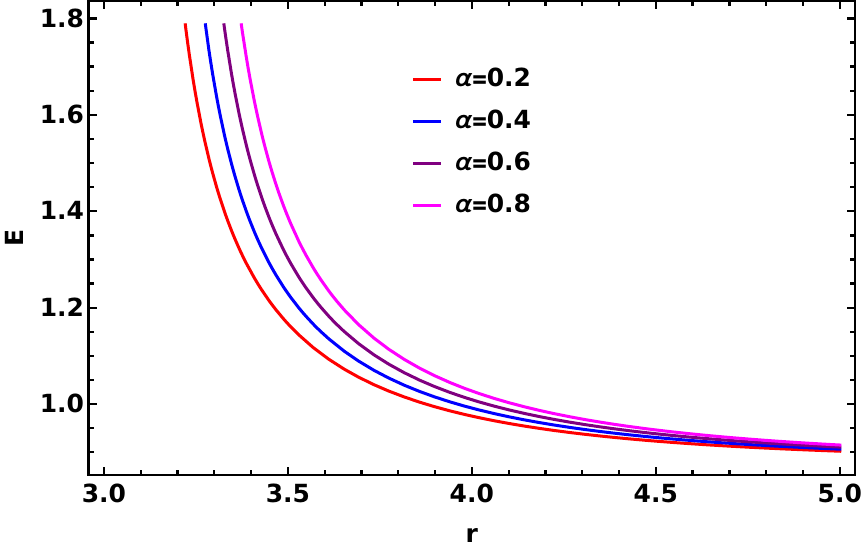}}\\
    \subfloat[$c=0.01,\alpha=0.5$]{\centering{}\includegraphics[width=0.4\linewidth]{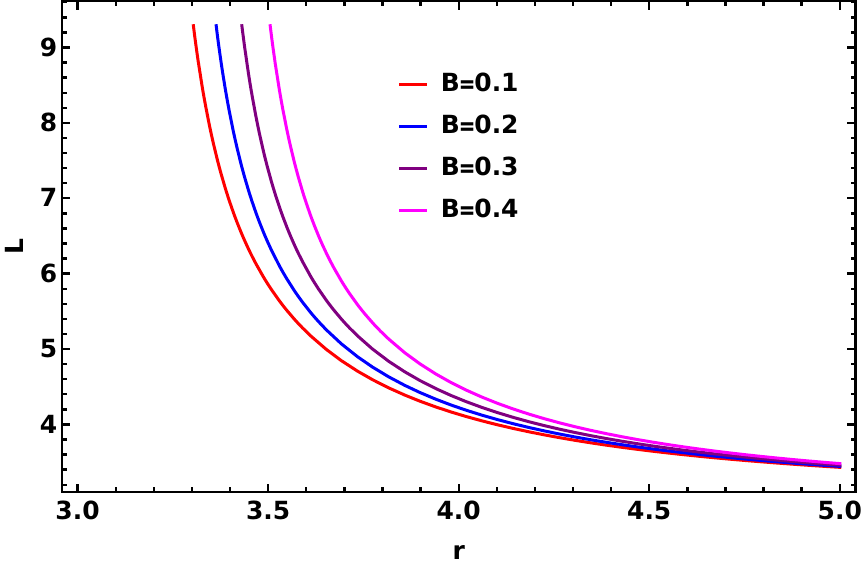}}\quad\quad\quad
    \subfloat[$c=0.01,\alpha=0.5$]{\centering{}\includegraphics[width=0.41\linewidth]{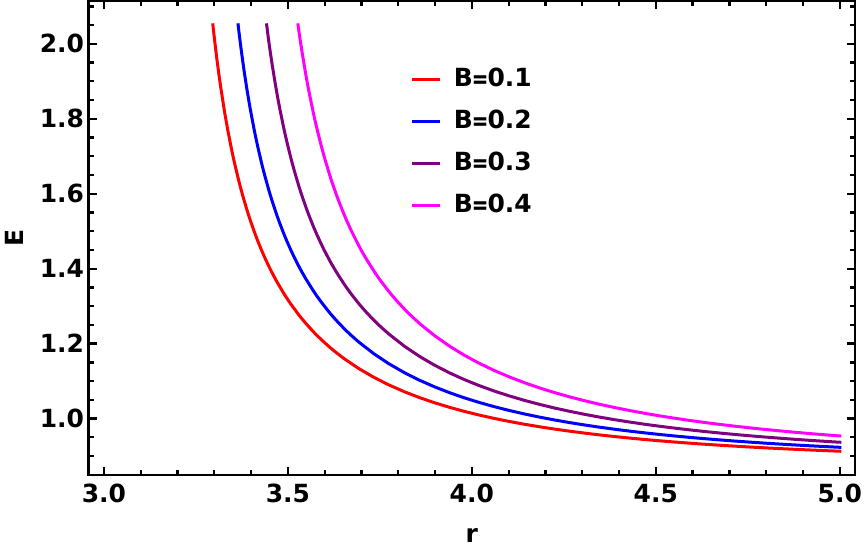}}\\
    \subfloat[$\alpha=0.5,\beta=0.1$]{\centering{}\includegraphics[width=0.4\linewidth]{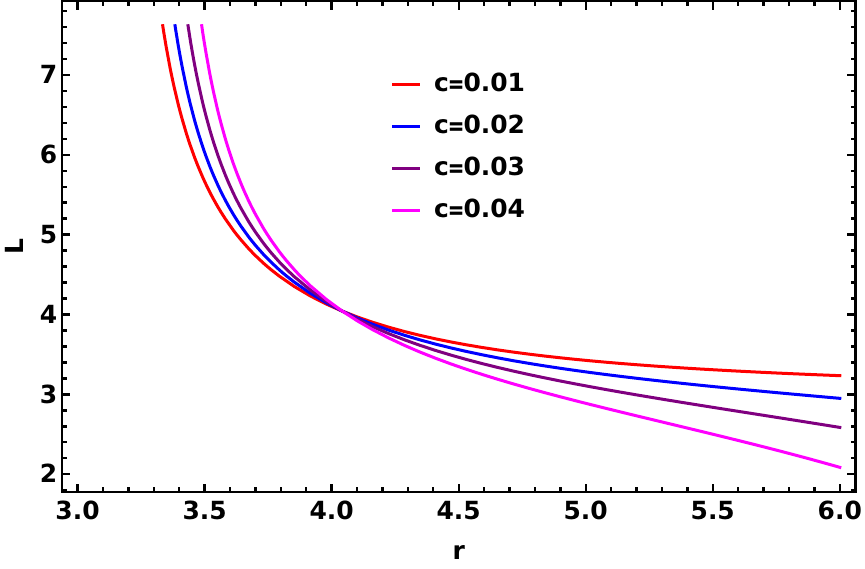}}\quad\quad\quad
    \subfloat[$\alpha=0.5,\beta=0.1$]{\centering{}\includegraphics[width=0.41\linewidth]{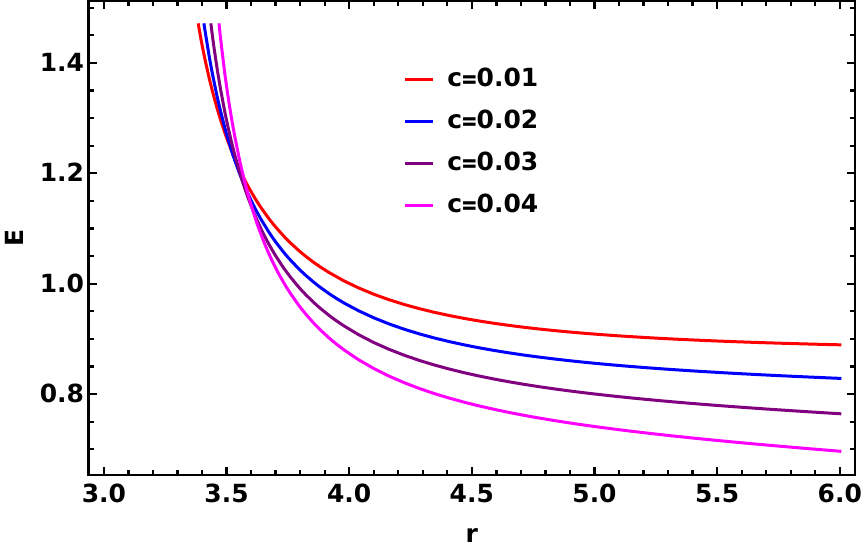}}\\
    \subfloat[]{\centering{}\includegraphics[width=0.4\linewidth]{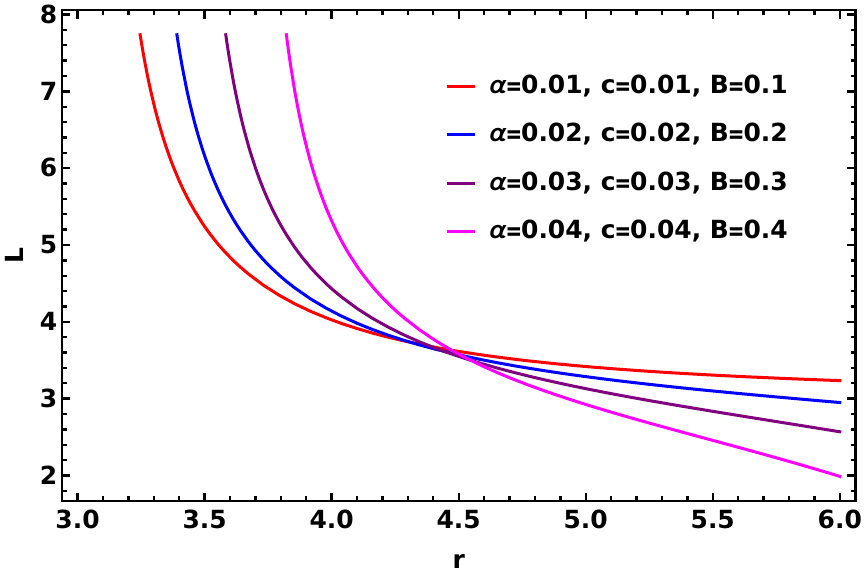}}\quad\quad\quad
    \subfloat[]{\centering{}\includegraphics[width=0.41\linewidth]{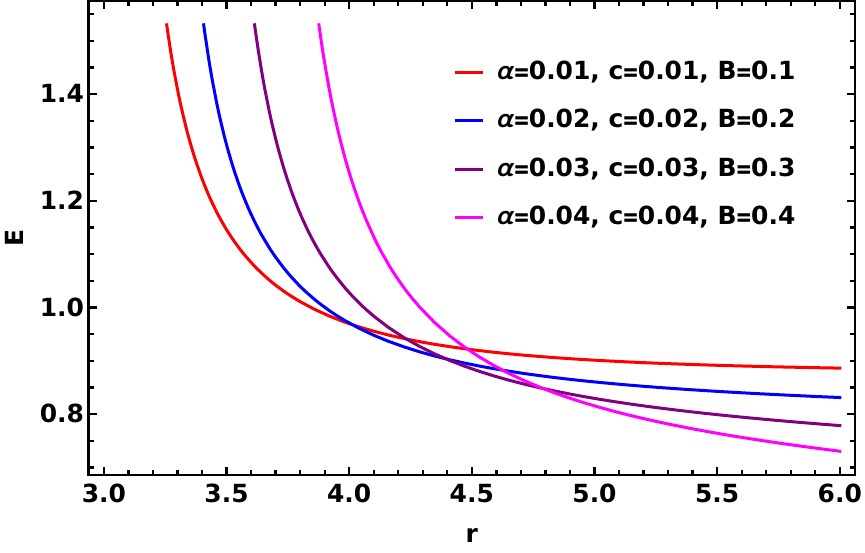}}
    \caption{The behavior of the specific angular momentum ($\mathrm{L}$) and energy ($\mathrm{E}$) of time-like particles for different values of $c, \alpha$ and $B$. Here, we set $M=1, w=-2/3$.}
    \label{fig:momentum-energy}
\end{figure}

For circular time-like orbits, employing the conditions $\dot{r}=0$ and $\ddot{r}=0$ yields the following conserved angular momentum given by
\begin{equation}
    \mathrm{L}(r)=r\,\sqrt{\frac{r\,f'(r)}{2\,f(r)-r\,f'(r)}}=r\,\sqrt{\frac{\frac{M}{r}+\frac{c\,(3\,w+1)}{2\,r^{3\,w+1}}-\frac{\alpha}{4\,r^2}\,\left(B+\frac{2\,M}{r}\right)\left(B+\frac{4\,M}{r}\right)}{1-\frac{3\,M}{r}-\frac{c\,(3\,w+3)}{2\,r^{3\,w+1}}+\frac{\alpha}{2\,r^2}\,\left(B+\frac{3\,M}{r}\right)\,\left(B+\frac{2\,M}{r}\right)}}.\label{dd3}
\end{equation}
And the particles' energy
\begin{equation}
    \mathrm{E}_{\pm}(r)=\pm\,\sqrt{\frac{2}{2\,f(r)-r\,f'(r)}}\,f(r)=\pm\,\frac{\left(1-\frac{2\,M}{r}-\frac{c}{r^{3\,w+1}}+\frac{\alpha}{r^2}\left(\frac{B}{2}+\frac{M}{r}\right)^2\right)}{\sqrt{1-\frac{3\,M}{r}-\frac{c\,(3\,w+3)}{2\,r^{3\,w+1}}+\frac{\alpha}{2\,r^2}\,\left(B+\frac{3\,M}{r}\right)\,\left(B+\frac{2\,M}{r}\right)}}.\label{dd4}
\end{equation}

From expressions given in Eqs. (\ref{dd3}) and (\ref{dd4}), it becomes evident that the angular momentum and energy associated with time-like particles orbiting in circular time-like geodesics is influenced by several factors. These include the coupling parameter $B$, the parameter $\alpha$, QF parameters $(c, w)$. Additionally, the BH mass $M$ also influence these physical quantities of time-like particles.

As for example, setting the state parameter $w=-2/3$, these quantities are given by the following expressions:
\begin{eqnarray}
    &&\mathrm{L}(r)=r\,\sqrt{\frac{\frac{M}{r}-\frac{c\,r}{2}-\frac{\alpha}{r^2}\,\left(\frac{B}{2}+\frac{M}{r}\right)\left(\frac{B}{2}+\frac{2\,M}{r}\right)}{1-\frac{3\,M}{r}-\frac{c\,r}{2}+\frac{\alpha}{2\,r^2}\,\left(B+\frac{3\,M}{r}\right)\,\left(B+\frac{2\,M}{r}\right)}},\label{dd5}\\ 
    &&\mathrm{E}_{\pm}(r)=\pm\,\frac{\left(1-\frac{2\,M}{r}-c\,r+\frac{\alpha}{r^2}\left(\frac{B}{2}+\frac{M}{r}\right)^2\right)}{\sqrt{1-\frac{3\,M}{r}-\frac{c\,r}{2}+\frac{\alpha}{2\,r^2}\,\left(B+\frac{3\,M}{r}\right)\,\left(B+\frac{2\,M}{r}\right)}}.\label{dd6}
\end{eqnarray}

In Figure \ref{fig:momentum-energy}, we present a series of plots illustrating the behavior of the specific angular momentum (left column) and specific energy (right column) as the parameters $\alpha$, $B$, $c$, and their combinations vary, while keeping the other parameters fixed along with the state parameter $w = -2/3$.

Now, we determine the orbital angular velocity of time-like particle and analyze how various factors involved in the space-time influence on it. This orbital is given by
\begin{equation}
    \Omega=\frac{\dot{\phi}}{\dot{t}}=\sqrt{\frac{f'(r)}{2\,r}}=\sqrt{\frac{M}{r^3}+\frac{c\,(3\,w+1)}{2\,r^{3\,w+3}}-\frac{\alpha}{r^4}\,\left(\frac{B}{2}+\frac{M}{r}\right)\,\left(\frac{B}{2}+\frac{2\,M}{r}\right)}.\label{dd7}
\end{equation}

From expression given in Eq. (\ref{dd7}), it becomes evident that the orbital angular velocity of time-like particles in circular orbits is influenced by several factors. These include the coupling parameter $B$, the parameter $\alpha$, QF parameters $(c, w)$. Additionally, the BH mass $M$ also influence these physical quantities of time-like particles.

As for example, setting the state parameter $w=-2/3$, the orbital angular velocity of time-like particles in circular orbits becomes
\begin{equation}
    \Omega=\sqrt{\frac{M}{r^3}-\frac{c}{2\,r}-\frac{\alpha}{r^4}\,\left(\frac{B}{2}+\frac{M}{r}\right)\,\left(\frac{B}{2}+\frac{2\,M}{r}\right)}.\label{dd8}
\end{equation}

Now, we aim to find speed with which timelike particle orbits the BH at a very large distance in comparison with the horizon of the BH. This is in analogy with a distant star in a galaxy moving in a circle around the BH of the galaxy. In the zeroth approximation, one write
\begin{equation}
    f(r)=1+2\,\Phi(r)\label{dd9}
\end{equation}
in which $\Phi(r)$ is the Newtonian gravitational potential for the time-like particle of unit mass. In our case, this gravitational potential is given by
\begin{equation}
    \Phi(r)=\frac{1}{2}\,\left(-\frac{2\,M}{r}-\frac{c}{r^{3\,w+1}}+\frac{\alpha}{r^2}\left(\frac{B}{2}+\frac{M}{r}\right)^2\right).\label{dd10}
\end{equation}

Therefore, the effective gravitational force is given by 
\begin{equation}
\mathrm{F}_c=-\frac{\partial \Phi(r)}{\partial r}\label{dd11}    
\end{equation}
which is toward the center of the BH. 
 
Substituting the gravitational potential given in eq. (\ref{dd10}) into the Eq. (\ref{dd11}), we find this effective force given by
\begin{eqnarray}
    \mathrm{F}_c=-\frac{M}{r^2}-\frac{c\,(3\,w+1)}{2\,r^{3\,w+2}}+\frac{\alpha}{r^3}\,\left(\frac{B}{2}+\frac{M}{r}\right)\,\left(\frac{B}{2}+\frac{2\,M}{r}\right).\label{dd12}
\end{eqnarray}
This central force can be equated with centripetal acceleration, {\it i.e.}, $|\mathrm{F}_c|=\frac{v^2}{r}$ in which $v$ is the speed of the test particle in the orbit. This equation results an expression for the circular speed $v$ given by
\begin{eqnarray}
    v=\sqrt{\Big|-\frac{M}{r}-\frac{c\,(3\,w+1)}{2\,r^{3\,w+1}}+\frac{\alpha}{r^2}\,\left(\frac{B}{2}+\frac{M}{r}\right)\,\left(\frac{B}{2}+\frac{2\,M}{r}\right)\Big|}.\label{dd13}
\end{eqnarray}
The expression (\ref{dd13}) shows that the speed of time-like particles at large distance from BH is influenced by the coupling parameter $B$, the parameter $\alpha$, QF parameters $(c, w)$, and the BH mass $M$.

For a state parameter $w=-2/3$, we find the speed
\begin{eqnarray}
    v=\sqrt{\Big|-\frac{M}{r}+\frac{c\,r}{2}+\frac{\alpha}{r^2}\,\left(\frac{B}{2}+\frac{M}{r}\right)\,\left(\frac{B}{2}+\frac{2\,M}{r}\right)\Big|}.\label{dd14}
\end{eqnarray}

\section {Shadow of LQG BH with QF } \label{isec4}

\begin{figure}[ht!]
    \centering
    \includegraphics[width=0.45\linewidth]{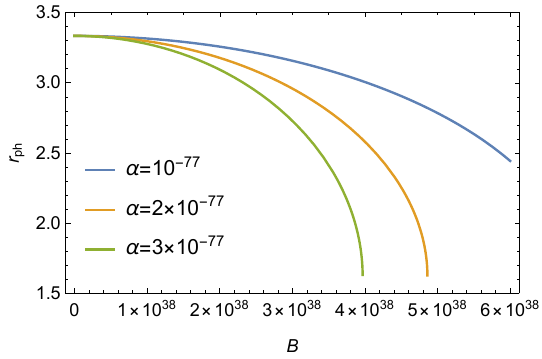}\quad\quad
    \includegraphics[width=0.45\linewidth]{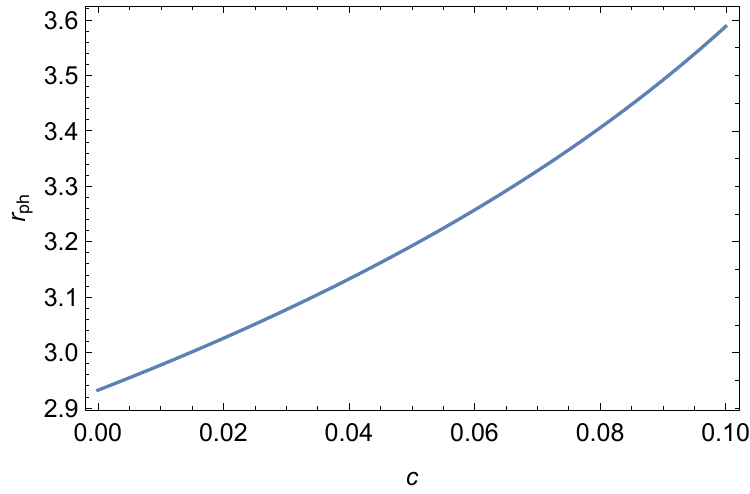}
    \caption{Variation of photon sphere radius for various values of BH parameters $\alpha$  and $B$ for fixed $c=0.06$ (left), for various values of $c$ for fixed $\alpha=10^{-77}$  and $B=2\times10^{38}$ (right).}
    \label{figa1}
\end{figure}

In this section, we will look at the shadow and shadow radius behaviour of a LQG BH surrounded by QF.   Several methods have been developed for calculating the shadow radius of a spherical BH. Ref. \cite{shad1} provides a current summary. For a static spherically symmetric metric, Eq. \cite{shad1} gives the radius of the photon ring around the BH.
\begin{equation}
    r\,D'(r)=D(r),
\end{equation}
where $D=\sqrt{f(r)}$. Considering  the metric function (\ref{aa2}) with especial choose $w=-2/3$, then the equation for photon sphere $r_\text{ph}$ is 
\begin{equation}
c\,r^5-(2r^4-6Mr^3+B^2\,\alpha \,r^2+5B\,M\alpha\,r+6M^2\,\alpha)=0. \label{eps1}
\end{equation} 
when $c=0=w$ and $\alpha=0$   then  equation (\ref{eps1}) reduces to $3M$. It is not possible to solve this equation (\ref{eps1}) analytically. Therefore, we use a numerical technique to get the photon orbit radii $r_{ph}$. In Figure \ref{figa1} we provide a comprehensive illustration of the influence of the BH parameters on photon sphere. The left panel of Figure \ref{figa1} represents the plot of $r_{ph}$ versus $B$ for different values of $\alpha$ and fixed $c=0.06$. It  clearly shows that the photon radius decreases with the parameters $\alpha, B$. The right panel of Figure \ref{figa1} represents the variation of $r_{ph}$ with the parameter $c$ by setting $\alpha$ and $B$ constant. It shows that the photon sphere increases with the parameter  $c$. We note that the effects of the QF parameter, $c$, and the parameters $\alpha,B$, on photon radii differ. 

After getting the photon orbit radii, we may calculate the shadow radii directly using \begin{equation}
    R_s=\frac{r_\text{ph}}{\sqrt{f(r_\text{ph})}} .\label{shadeq1}
\end{equation}
Table \ref{taba3} shows the numerical values of  photon sphere $r_{ph}$ and shadow radii $R_s$ for various parameters of LQG BH with QF.

{\small
\begin{center}
\begin{tabular}{|c|c c|c c|c c|cc|}
 \hline  \rowcolor{lightgray} & \multicolumn{2}{|c|}{ $B=1\times 10^{38}$} & \multicolumn{2}{|c|}{ $2\times 10^{38}$} &  \multicolumn{2}{|c|}{ $3\times 10^{38}$}&  \multicolumn{2}{|c|}{ $4\times 10^{38}$} \\ \hline  \rowcolor{lightgray}
$\alpha $ & $r_{ph}$ & $R_{s}$ & $r_{ph}$ & $R_{s}$ & $r_{ph}$ & $R_{s}$ & $%
r_{ph}$ & $R_{s}$ \\ \hline
$1\times 10^{-77}$ & $3.31449$ & $7.41151$ & $3.25679$ & $7.28388$ & $3.15629
$ & $7.06558$ & $3.00456$ & $6.78326$ \\ 
$2\times 10^{-77}$ & $3.29546$ & $7.36922$ & $3.17686$ & $7.10985$ & $2.95832
$ & $6.65068$ & $2.57923$ & $5.91595$ \\ 
$3\times 10^{-77}$ & $3.27623$ & $7.32668$ & $3.09302$ & $6.93077$ & $2.7286$
& $6.19546$ & $1.92861$ & $4.91522$\\
  \hline
\end{tabular}
\captionof{table}{Numerical results for the photon radius and shadow radius with various BH parameters, $\alpha$ and  $B$. Here $c=0.06$ and $w=-2/3$.} \label{taba3}
\end{center}
}
Table \ref{taba3} shows that the effects of the parameter, $\alpha, B$ on shadow radii. Increasing  the parameters $\alpha, B$ lead to a decrease in shadow radius.

\begin{figure}[ht!]
    \centering
    \includegraphics[width=0.45\linewidth]{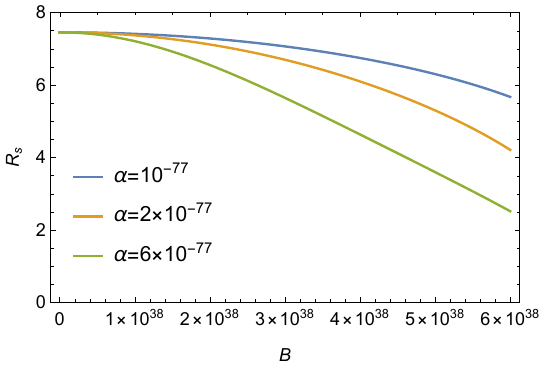}\quad\quad
    \includegraphics[width=0.45\linewidth]{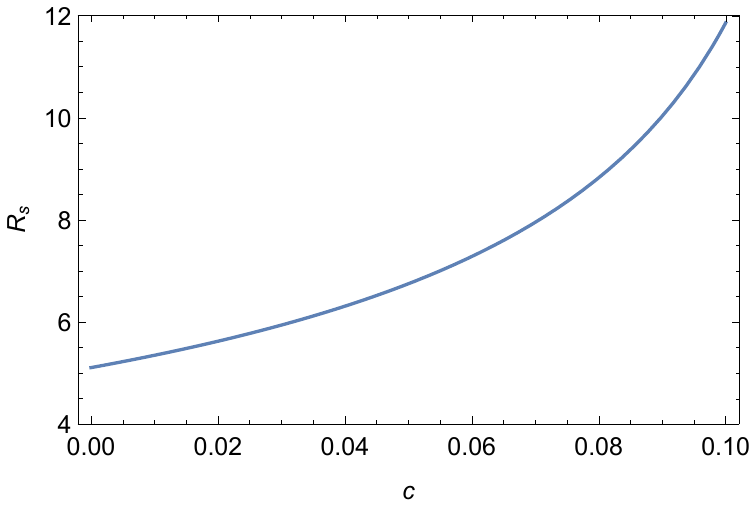}
    \caption{Variation of shadow radius for various values of BH parameters $\alpha$  and $B$ for fixed $c=0.06$ (left), for various values of $c$ for fixed $\alpha=10^{-77}$  and $B=2\times10^{38}$ (right).}
    \label{figa2}
\end{figure}

  Figure  \ref{figa2} provides a comprehensive illustration of BH parameters behaviors.
\\To represent the actual shadow of the BH as seen from an observer's perspective, we introduce celestial coordinates, $X$ and $Y$
\begin{equation}
X=\lim_{r_{\mathrm{o}}\rightarrow \infty }\left( -r_{\mathrm{o}}^{2}\sin
\theta _{\mathrm{o}}\frac{d\varphi }{dr}\right) ,
\end{equation}%
\begin{equation}
Y=\lim_{r_{\mathrm{o}}\rightarrow \infty }\left( r_{\mathrm{o}}^{2}\frac{%
d\theta }{dr}\right) .
\end{equation}

\begin{figure}[ht!]
    \centering
    \includegraphics[scale=0.45]{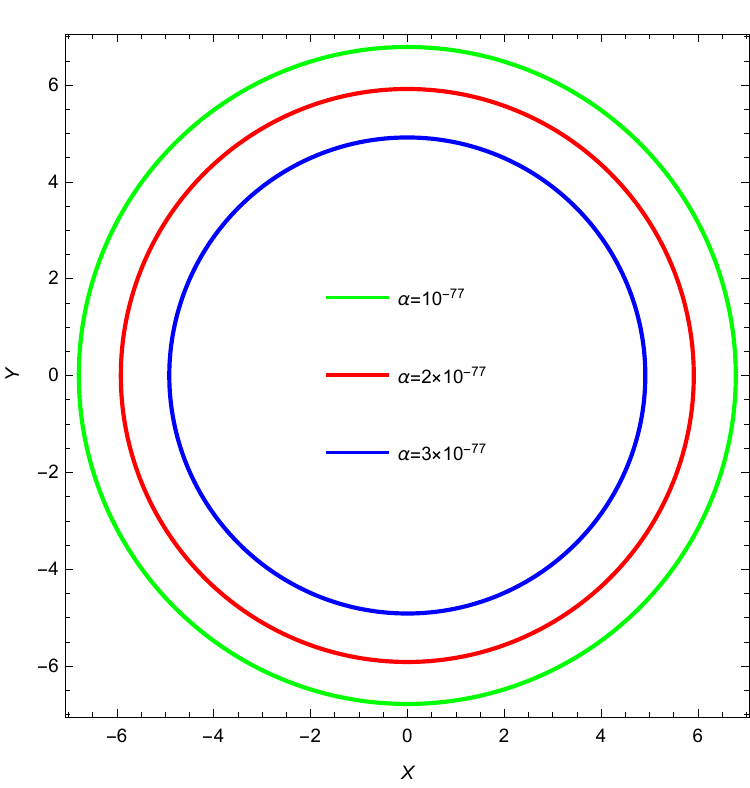} \includegraphics[scale=0.45]{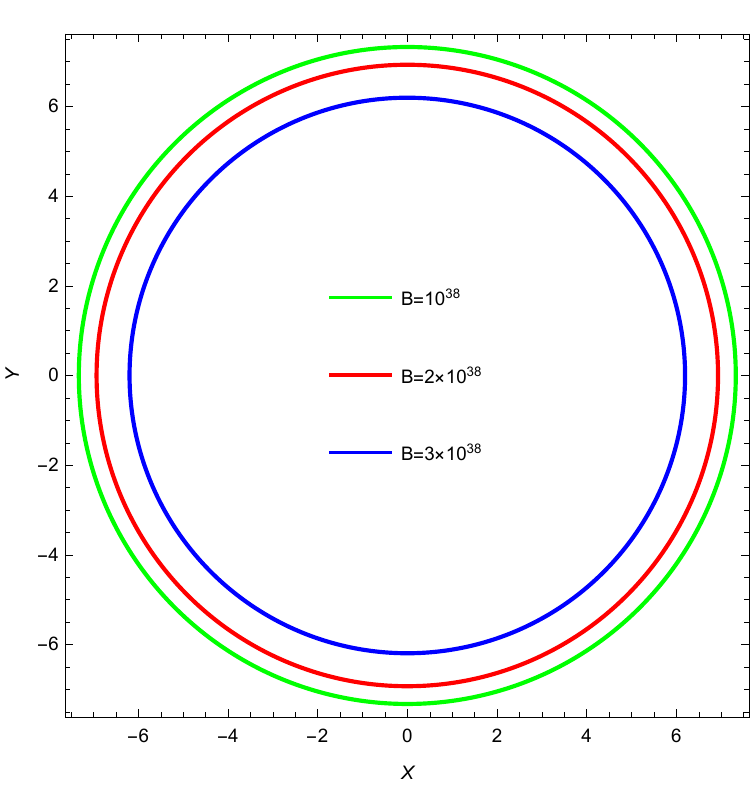}
    \includegraphics[scale=0.45]{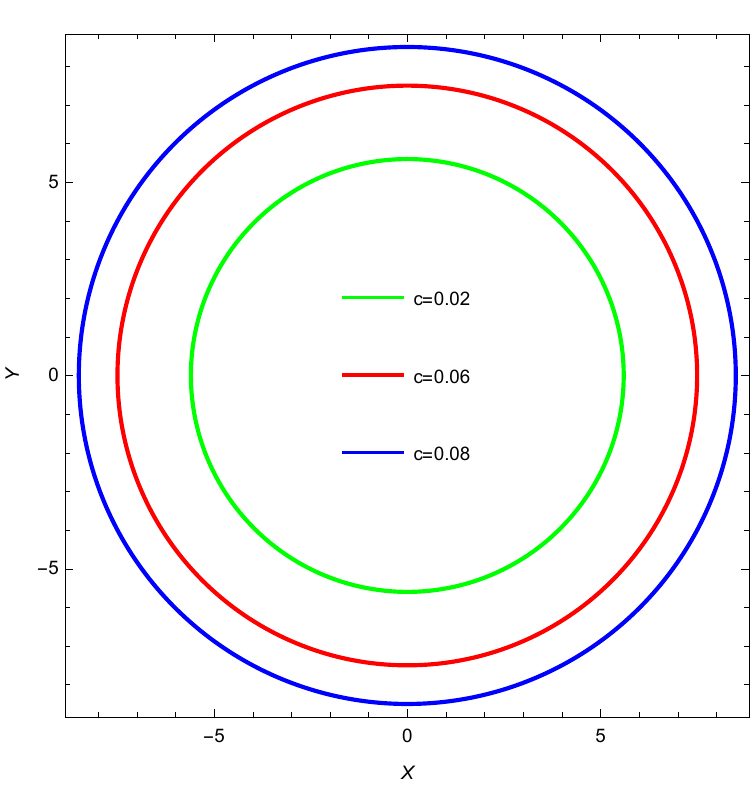}
    \caption{BH shadows  for different values of $\alpha$ (left) of $B$ (middle) and for different values of $c$ (right).}
    \label{ps25}
\end{figure}

For a static observer at large distance, i.e. at  $r_{\mathrm{o}}\rightarrow
\infty $ in the equatorial plane $\theta _{\mathrm{o}}=\pi /2$, the
celestial coordinates simplify to 
\begin{equation}
X^{2}+Y^{2}=R_s^{2}.
\end{equation}
Figure \ref{ps25} shows how the shadow varies with the BH parameters. It shows that the presence of quintessence causes the shadow size to increase while decreasing with the parameters $\alpha$ and $B$.

\section{Scalar Perturbations} \label{isec5}

In this section, we investigate the dynamics of a massless scalar field in the background of a selected deformed BH solution with a global monopole charge under thermal fluctuations. We begin by explicitly deriving the massless Klein-Gordon equation, which governs the evolution of the scalar field in the given spacetime geometry. 

Scalar perturbations, in the context of BH space-times, are crucial for understanding BH stability. This perturbation has been widely studied in various BH solutions in GR, where they provide important insights into the nature of BH stability and the propagation of fields in curved spacetime. For instance, scalar field perturbations have been analyzed in Schwarzschild, Kerr, and Reissner-Nordström BHs, along with other BHs solutions in GR and modified gravity theories (see {\it e.g.}, Refs. \cite{NPB2,CJPHY,EPJC2,PDU1,PDU2,NPB3}, and other related works).

The massless scalar field wave equation is described by the Klein-Gordon equation:
\begin{equation}
\frac{1}{\sqrt{-g}}\,\partial_{\mu}\left[\left(\sqrt{-g}\,g^{\mu\nu}\,\partial_{\nu}\right)\,\Psi\right]=0,\label{ff1}    
\end{equation}
where $\Psi$ is the wave function of the scalar field, $g_{\mu\nu}$ is the covariant metric tensor, $g=\det(g_{\mu\nu})$ is the determinant of the metric tensor, $g^{\mu\nu}$ is the contrvariant form of the metric tensor, and $\partial_{\mu}$ is the partial derivative with respect to the coordinate systems.

Before, writing explicitly, performing the following coordinate change (called tortoise coordinate) 
\begin{eqnarray}
    dr_*=\frac{dr}{f(r)}\label{ff2}
\end{eqnarray}
into the line-element Eq. (\ref{bb1}) results
\begin{equation}
    ds^2=f(r_*)\,\left(-dt^2+dr^2_{*}\right)+h^2(r_*)\,\left(d\theta^2+\sin^2 \theta\,d\phi^2\right),\label{ff3}
\end{equation}
where $f(r_*)$ and $h(r_*)$ are functions of $r_*$. 

Let us consider the following scalar field wave function ansatz form
\begin{equation}
    \Psi(t, r_{*},\theta, \phi)=\exp(i\,\omega\,t)\,Y^{m}_{\ell} (\theta,\phi)\,\frac{\psi(r_*)}{r_{*}},\label{ff4}
\end{equation}
where $\omega$ is (possibly complex) the temporal frequency, $\psi (r)$ is a propagating scalar field in the candidate spacetime, and $Y^{m}_{\ell} (\theta,\phi)$ is the spherical harmonics.

With these, we can write the wave equation (\ref{ff1}) in the following form:
\begin{equation}
    \frac{\partial^2 \psi(r_*)}{\partial r^2_{*}}+\left(\omega^2-\mathcal{V}\right)\,\psi(r_*)=0,\label{ff5}
\end{equation}
where the scalar perturbative potential $\mathcal{V}$ is given by 
\begin{eqnarray} 
\mathcal{V}&=&\left(\frac{\ell\,(\ell+1)}{r^2}+\frac{f'(r)}{r}\right)\,f(r).\label{ff6}
\end{eqnarray}

Substituting the metric function $f(r)$ we find the scalar perturbative potential given by
\begin{eqnarray}
    \mathcal{V}=\left[\frac{\ell\,(\ell+1)}{r^2}+\frac{2\,M}{r^3}+\frac{c\,(3\,w+1)}{r^{3\,w+3}}-\frac{2\,\alpha}{r^4}\,\left(\frac{B}{2}+\frac{M}{r}\right)\,\left(\frac{B}{2}+\frac{2\,M}{r}\right)\right]\left[1-\frac{2\,M}{r}-\frac{c}{r^{3\,w+1}}+\frac{\alpha}{r^2}\left( \frac{B}{2}+\frac{M}{r}\right)^2\right].\label{ff7}
\end{eqnarray}

From expression given in Eq. (\ref{ff8}), it becomes evident that the perturbative potential is influenced by several factors. These include the coupling parameter $B$, the parameter $\alpha$, QF parameters $(c, w)$. Additionally, the BH mass $M$ also influence this potential.

\begin{figure}[ht!]
    \centering
    \subfloat[$c=0.01, B=0.5$]{\centering{}\includegraphics[width=0.45\linewidth]{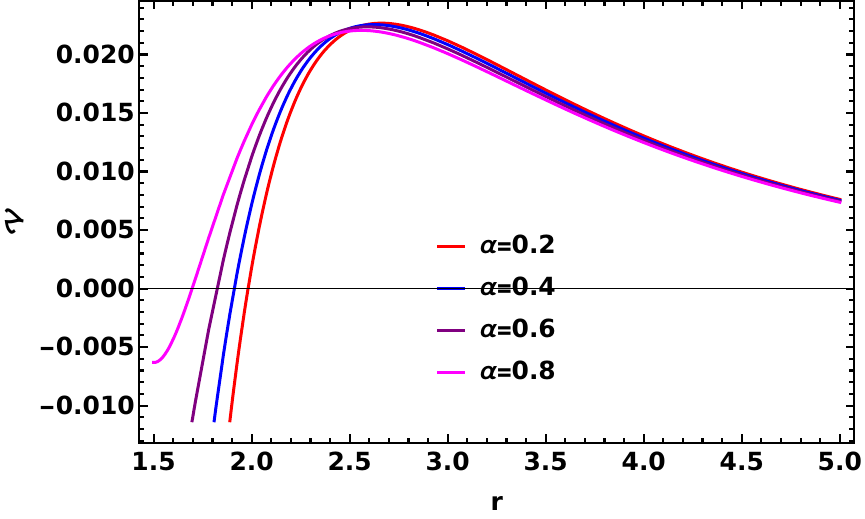}}\quad\quad
    \subfloat[$c=0.01, \alpha=0.5$]{\centering{}\includegraphics[width=0.45\linewidth]{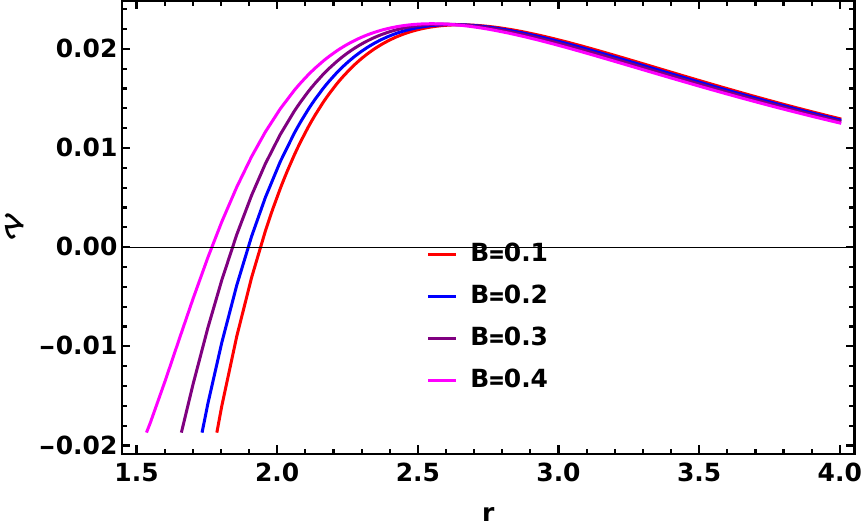}}\\
    \subfloat[$\alpha=0.5, B=0.1$]{\centering{}\includegraphics[width=0.45\linewidth]{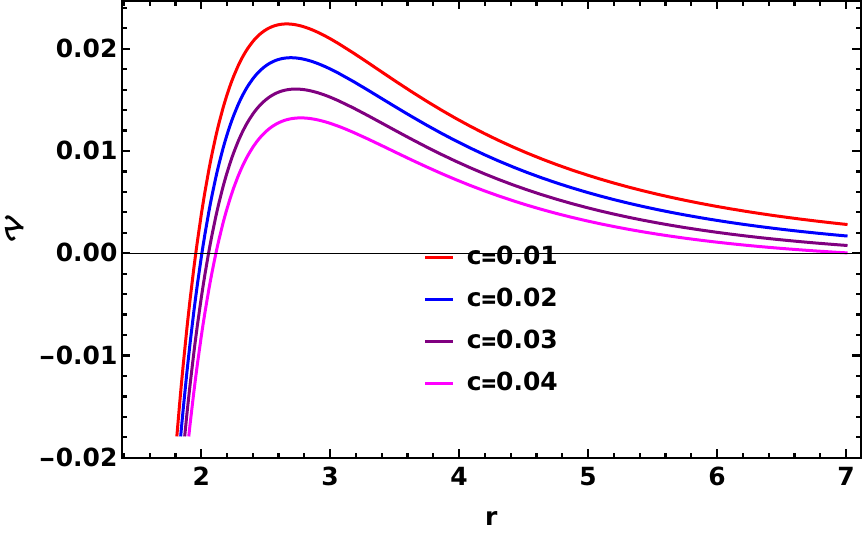}}\quad\quad
    \subfloat[]{\centering{}\includegraphics[width=0.45\linewidth]{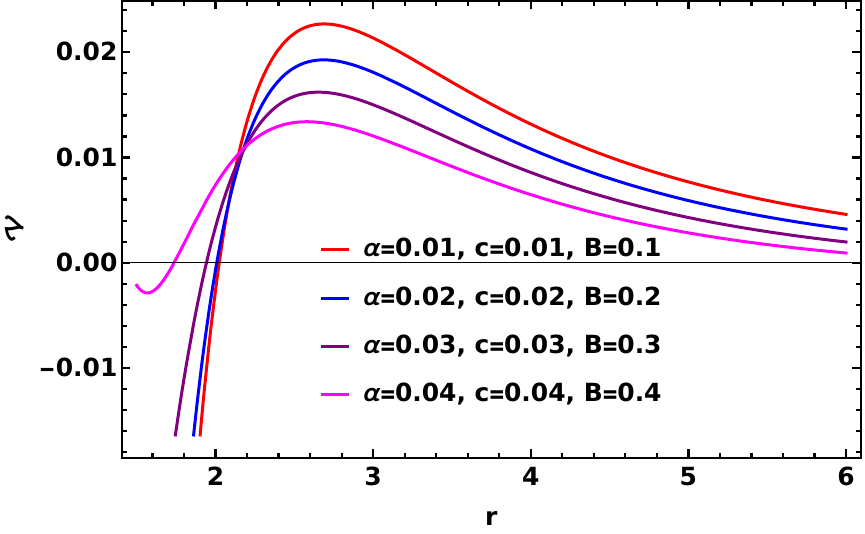}}
    \caption{The behavior of the scalar perturbative potential as a function of $r$ for different values of $\alpha, B$ and $c$. Here, we set $M=1, \ell=0, w=-2/3$.}
    \label{fig:scalar-potential}
\end{figure}

As for example, setting the state parameter $w=-2/3$, we find the potential
\begin{eqnarray}
    \mathcal{V}=\left[\frac{\ell\,(\ell+1)}{r^2}+\frac{2\,M}{r^3}-\frac{c}{r}-\frac{2\,\alpha}{r^4}\,\left(\frac{B}{2}+\frac{M}{r}\right)\,\left(\frac{B}{2}+\frac{2\,M}{r}\right)\right]\left[1-\frac{2\,M}{r}-c\,r+\frac{\alpha}{r^2}\left( \frac{B}{2}+\frac{M}{r}\right)^2\right].\label{ff8}
\end{eqnarray}

In Figure \ref{fig:scalar-potential}, we present a series of plots illustrating the behavior of the scalar perturbative potential $\mathcal{V}$ for $\ell=0$-state (called s-state) as the parameters $\alpha$, $B$, $c$, and their combinations vary, while keeping the other parameters fixed along with the state parameter $w = -2/3$.

\subsection{Quasinormal Modes (QNMs)} 

In this part, we will look at the scalar and electromagnetic field perturbations of the LQG BH in the context of the quintessence field to see how QNMs behave. Gravitational waves released by BH coalescence provide significant information about the nature of spacetime and are unaffected by any particular beginning stage. They also reveal information on the BH system's stability in the face of perturbations. QNMs for scalar and electromagnetic field perturbations are wave equation solutions that meet specific boundary conditions both near and far from the BH horizon. The solution must satisfy the criteria for purely ingoing waves at the event horizon and purely outgoing waves at the cosmological horizon or spatial infinity.\\ The WKB approximation method is often used to calculate QNMs. Iyer \cite{d3} introduced it first, and Konoplya \cite{d4} expanded it to higher levels subsequently. The WKB technique is effective for low overtone values $n$, especially when $n <l$.  Using the effective potentials obtained in the preceding section, we numerically calculate the quasinormal frequencies for scalar and electromagnetic perturbations using the 6th order WKB approximation.\\ The calculated QNM frequencies $\omega$ are presented in Tables \ref{taba13}, \ref{taba14} and \ref{taba15}, showing their dependence on the parameters $\alpha$, $c$, and $B$, respectively. 

\begin{center}
\begin{tabular}{|c|c|c|}
 \hline 
 \rowcolor{gray!50}
 \multicolumn{3}{|c|}{ $c=0.02$, $B_0=0.4$, $n=0$, $\ell=2$}
\\ \rowcolor{gray!50} \hline $\alpha $ & $\mbox{Scalar}$ & $\mbox{EM}$ \\ \hline
$0$ & $0.493776-0.0941709i$ & $0.468161-0.0924586i$ \\ 
$0.01$ & $0.470257-0.0887972i$ & $0.447209-0.0871102i$ \\ 
$0.02$ & $0.446079-0.0833173i$ & $0.42555-0.0816705i$ \\ 
$0.03$ & $0.421142-0.0777194i$ & $0.403081-0.0761292i$ \\ 
$0.04$ & $0.395322-0.0719887i$ & $0.379671-0.070473i$ \\ 
$0.05$ & $0.368454-0.0661058i$ & $0.355147-0.0646844i$ \\ 
$0.06$ & $0.340317-0.0600449i$ & $0.329277-0.0587394i$ \\ 
$0.07$ & $0.310598-0.0537704i$ & $0.301737-0.0526044i$ \\ 
$0.08$ & $0.278834-0.0472284i$ & $0.272042-0.0462284i$
\\ 
 \hline
\end{tabular}
\captionof{table}{Variation of amplitude and damping of QNMs with respect to the Planck length parameter.} \label{taba13}
\end{center} 

\begin{center}
\begin{tabular}{|c|c|c|}
 \hline \rowcolor{gray!50}
 \multicolumn{3}{|c|}{ $\alpha=0.4$, $B_0=0.4$, $n=0$, $\ell=2$}
\\  \rowcolor{gray!50} \hline $c $ & $\mbox{Scalar}$ & $\mbox{EM}$ \\ \hline
$0$ & $0.436613-0.0852419i$ & $0.415851-0.0835375i$ \\ 
$0.1$ & $0.43885-0.0848496i$ & $0.418134-0.0831623i$ \\ 
$0.2$ & $0.44117-0.0844043i$ & $0.420507-0.0827329i$ \\ 
$0.3$ & $0.443577-0.0838972i$ & $0.422976-0.0822396i$ \\ 
$0.4$ & $0.446079-0.0833173i$ & $0.42555-0.0816705i$ \\ 
$0.5$ & $0.44868-0.082651i$ & $0.428236-0.0810106i$ \\ 
$0.6$ & $0.451388-0.0818814i$ & $0.431041-0.0802405i$ \\ 
$0.7$ & $0.454206-0.080987i$ & $0.433973-0.0793359i$ \\ 
$0.8$ & $0.457135-0.0799413i$ & $0.437035-0.078265i$
\\ 
 \hline
\end{tabular}
\captionof{table}{Variation of amplitude and damping of QNMs with respect to the normalization QF constant.} \label{taba14}
\end{center} 

\begin{center}
\begin{tabular}{|c|c|c|}
 \hline \rowcolor{gray!50} 
 \multicolumn{3}{|c|}{ $\alpha=0.4$, $c=0.02$, $n=0$, $\ell=2$}
\\ \rowcolor{gray!50} \hline $B_0 $ & $\mbox{Scalar}$ & $\mbox{EM}$ \\ \hline
$0$ & $0.440119-0.0839573i$ & $0.419534-0.0822767i$ \\ 
$0.2$ & $0.442697-0.0836503i$ & $0.422142-0.0819862i$ \\ 
$0.4$ & $0.446079-0.0833173i$ & $0.42555-0.0816705i$ \\ 
$0.6$ & $0.450358-0.0829233i$ & $0.429857-0.0812938i$ \\ 
$0.8$ & $0.455661-0.0824141i$ & $0.435197-0.0807996i$ \\ 
$1$ & $0.462164-0.081701i$ & $0.441756-0.080094i$
\\ 
 \hline
\end{tabular}
\captionof{table}{Variation of amplitude and damping of QNMs with respect to the coupling parameter.} \label{taba15}
\end{center} 

The quasinormal frequencies listed in Tables \ref{taba13}-\ref{taba15} include both a real part, corresponding to the oscillation frequency, and an imaginary part, which determines the damping rate. The negative sign of the imaginary component signifies the stability of the perturbations, as it ensures an exponential decay of the oscillations over time rather than uncontrolled growth. This observation confirms that the LQG BH surrounded by a QF remains stable under scalar and electromagnetic perturbations, at least within the parameter ranges examined in this study.

\begin{figure}[ht!]
    \centering
    \includegraphics[width=0.9\linewidth]{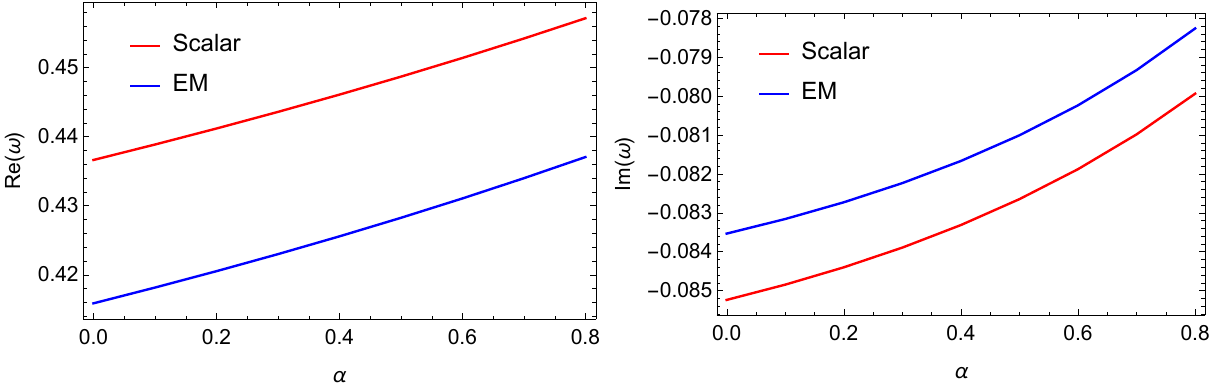}
    \caption{ Variation of amplitude and damping of QNMs with respect to the  Planck length parameter  for scalar and EM perturbations.}
    \label{realRo}
    \includegraphics[width=0.9\linewidth]{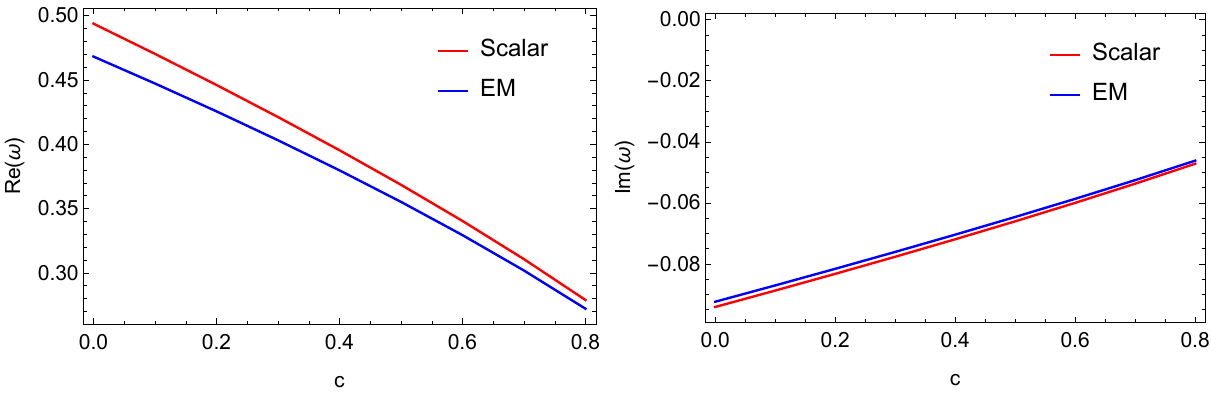}
    \caption{ Variation of amplitude and damping of QNMs with respect to the normalization QF constant  for scalar and EM perturbations.}
    \label{realS}
    \includegraphics[width=0.9\linewidth]{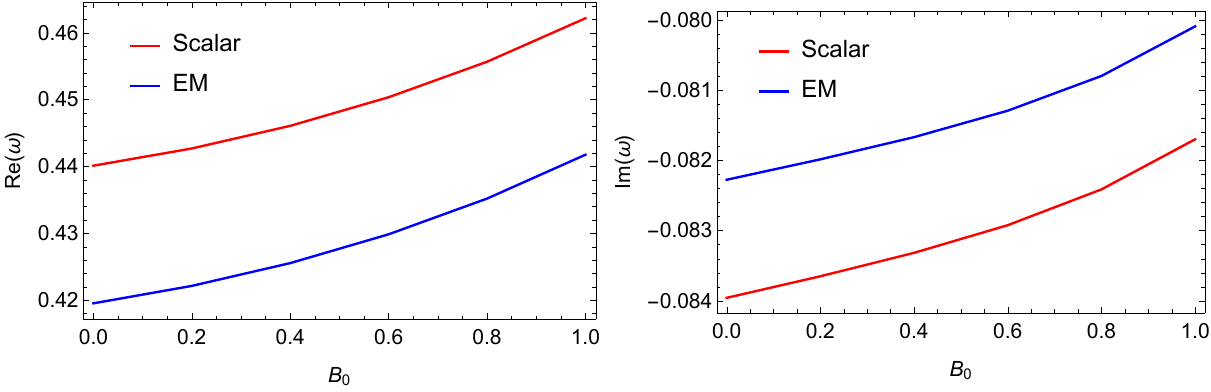}
    \caption{ Variation of amplitude and damping of QNMs with respect to the coupling parameter  for scalar and EM perturbations.}
    \label{realG}
\end{figure}

In figure \ref{realRo}, we illustrate the variation of real and imaginary QNMs with regard to the model parameter $\alpha$ for
massless scalar perturbation and electromagnetic perturbation, it is shown that the real part of the QNM frequency, $Re(\omega)$, increases with the Planck length parameter $\alpha$. This suggests that the oscillation frequency of the perturbations steadily increases, resulting in a greater reaction of the BH spacetime. As $\alpha$ increases, the imaginary part $Im(\omega)$ becomes less negative. This means that the disturbances dissipate at a slower rate, meaning that the oscillations would last longer (see figure \ref{realRo}). A similar pattern is seen with the coupling parameter parameter $B$ as shown in figure  \ref{realG}.

The study also investigates the QNM frequencies in relation to the normalization constant of QF $c$. In figure \ref{realS}, we illustrate the variation of real and imaginary QNMs with regard to the model parameter $c$ for massless scalar perturbation and electromagnetic perturbation.  As $c$ grows, $Re(\omega)$ decreases, indicating fewer stable oscillations. As $c$ increases, $Im(\omega)$  becomes less negative.\\ In summery, as $c$ increases, the real and imaginary parts of frequency $\omega$ fall monotonically, indicating that a strong QF leads to lower oscillation frequencies and longer-lasting disruptions. Combining LQG BH with QF affects the stability and dissipative behaviour of BH perturbations.

\section{Deflection Angle of LQG BHs via GBTm} \label{isec6}
{\color{black}

In this section, we investigate gravitational lensing by LQG BHs immersed in QFs. Recent research has emphasized that gravitational lensing is among the most reliable methods for examining the spacetime geometry surrounding compact objects \cite{ismz01,ismz02}. The interplay between quantum gravity corrections and exotic matter fields produces distinctive lensing signatures that may be detectable with current and future observational facilities \cite{ismz03,ismz04}.

To analyze these signatures, we employ the GBTm approach pioneered by Gibbons and Werner \cite{ismz05}, which has emerged as a powerful tool for studying deflection angles in modified gravity theories \cite{ismz06,ismz07,ijgmmp1,ijgmmp2,jcap}. This method offers significant computational advantages over traditional integration techniques while providing geometric insights into how spacetime deformations affect light propagation.

We begin our analysis within the weak-field approximation framework, considering a spherically symmetric and static BH solution immersed in a QF environment. The GBTm establishes a profound connection between the intrinsic curvature of spacetime and the global topological characteristics of a compact region $\mathcal{S}_{\mathcal{R}}$ bounded by $\partial \mathcal{S}_{\mathcal{R}}$. This fundamental relationship is expressed through the following mathematical formulation \cite{ism01,ism02}:
\begin{equation}\label{eq:gbt_main}
\iint_{\mathcal{S}_{R}} K dS + \oint_{\partial \mathcal{S}_{\mathcal{R}}} \kappa_g ds + \sum_{i=1}^{n} \epsilon_i = 2\pi\chi(\mathcal{S}_{\mathcal{R}}),
\end{equation}
where $\mathcal{S}_{\mathcal{R}} \subset \mathcal{M}_{2D}$ represents a compact domain within a two-dimensional differentiable manifold $\mathcal{M}_{2D}$, bounded by a smooth and oriented contour $\partial \mathcal{S}_{\mathcal{R}}$. Here, $K$ denotes the Gaussian optical curvature and $\kappa_g$ represents the geodesic curvature of $\partial \mathcal{S}_{\mathcal{R}}$, defined explicitly as $\kappa_g = g \left( \nabla_{\dot{\gamma}} \dot{\gamma}, \ddot{\gamma} \right)$, where $g(\dot{\gamma}, \dot{\gamma}) = 1$ and $\ddot{\gamma}$ is the unit acceleration vector. The term $\epsilon_i$ captures the exterior angle at the $i^{\text{th}}$ vertex of the boundary, while $\chi(\mathcal{S}_{\mathcal{R}})$ corresponds to the Euler characteristic number of the surface \cite{ism03}.

To employ the GBTm effectively, we must first determine the Gaussian optical curvature $K$ by examining null geodesics influenced by the BH surrounded by QF \cite{ism04,ism05}. Since light propagates along null geodesics (where $ds^2=0$), these paths naturally define an optical metric describing the effective Riemannian geometry experienced by light rays. By applying the null condition ($ds^2=0$) and restricting motion to the equatorial plane ($\theta = \pi/2$), where the optical metric provides a natural framework for rotational symmetry, we introduce the tortoise coordinate $r_*$ defined by:
\begin{equation}\label{eq:tortoise}
dr_* = \frac{dr}{f(r)},
\end{equation}
which allows us to express the optical metric in the transformed coordinate system:
\begin{equation}\label{eq:optical_metric}
dt^2 = g_{ij}^{\text{opt}}dx^i dx^j = dr_*^2 + H^2(r_*)d\phi^2,
\end{equation}
where 
\begin{equation}\label{eq:H_function}
H(r_*(r)) = \frac{r}{\sqrt{f(r)}}.
\end{equation}

To determine the Gaussian optical curvature $K$, we need to calculate the Christoffel symbols associated with the optical metric. The non-zero components are:
\begin{eqnarray}
\Gamma_{r_* \phi}^{\phi} &=& \frac{1}{H(r_*)}\frac{dH(r_*)}{dr_*}, \label{eq:christ1} \\
\Gamma_{\phi \phi}^{r_*} &=& -H(r_*)\frac{dH(r_*)}{dr_*}. \label{eq:christ2}
\end{eqnarray}

From these Christoffel symbols, we compute the Riemann curvature tensor component:
\begin{equation}\label{eq:riemann}
R_{r_* \phi r_* \phi} = \frac{\partial \Gamma_{\phi \phi}^{r_*}}{\partial r_*} - \frac{\partial \Gamma_{r_* \phi}^{r_*}}{\partial \phi} + \Gamma_{r_* \alpha}^{r_*}\Gamma_{\phi \phi}^{\alpha} - \Gamma_{\phi \alpha}^{r_*}\Gamma_{r_* \phi}^{\alpha} = -H(r_*)\frac{d^2H(r_*)}{dr_*^2}.
\end{equation}

The Gaussian optical curvature is then determined by:
\begin{equation}\label{eq:gaussian_curvature_1}
K = \frac{R_{r_* \phi r_* \phi}}{g_{r_* r_*}g_{\phi \phi} - g_{r_* \phi}^2} = \frac{R_{r_* \phi r_* \phi}}{H^2(r_*)} = -\frac{1}{H(r_*)}\frac{d^2H(r_*)}{dr_*^2}.
\end{equation}

Since our calculations involve the radial coordinate $r$, we need to express the Gaussian curvature in terms of $r$ rather than $r_*$. Using the chain rule and the relation between $r$ and $r_*$, we obtain:
\begin{eqnarray}
\frac{dH(r_*)}{dr_*} &=& \frac{dH(r)}{dr}\frac{dr}{dr_*} = \frac{dH(r)}{dr}f(r), \label{eq:chain1} \\
\frac{d^2H(r_*)}{dr_*^2} &=& \frac{d}{dr_*}\left(\frac{dH(r)}{dr}f(r)\right) = \frac{d}{dr}\left(\frac{dH(r)}{dr}f(r)\right)\frac{dr}{dr_*}
=\left[\frac{d^2H(r)}{dr^2}f(r) + \frac{dH(r)}{dr}\frac{df(r)}{dr}\right]f(r). \label{eq:chain2}
\end{eqnarray}

Substituting these expressions into Eq. (\ref{eq:gaussian_curvature_1}), we derive:
\begin{equation}\label{eq:gaussian_curvature_2}
K = -\frac{1}{H(r)}\left[f(r)^2\frac{d^2H(r)}{dr^2} + f(r)\frac{df(r)}{dr}\frac{dH(r)}{dr}\right].
\end{equation}

Further simplification using the explicit form of $H(r)$ from Eq. (\ref{eq:H_function}) yields:
\begin{equation}\label{eq:gaussian_curvature_3}
K = -\frac{f(r)}{r}\left[\frac{d^2}{dr^2}\left(\frac{r}{\sqrt{f(r)}}\right) + \frac{1}{f(r)}\frac{df(r)}{dr}\frac{d}{dr}\left(\frac{r}{\sqrt{f(r)}}\right)\right].
\end{equation}

After substituting the metric function from Eq. (\ref{aa2}) into Eq. (\ref{eq:gaussian_curvature_3}) and performing a series expansion for the weak-field approximation, we obtain the explicit form of the Gaussian optical curvature:
\begin{eqnarray}\label{eq:gaussian_curvature_expanded}
K &\approx& \frac{2M}{r^3} - \frac{c(3w+1)}{r^{3w+3}} + \frac{\alpha B^2}{2r^4} + \frac{2\alpha BM}{r^5} + \frac{3\alpha M^2}{r^6} + \mathcal{O}\left(\frac{1}{r^7}\right),
\end{eqnarray}

The Gaussian optical curvature expression reveals distinct physical contributions across multiple scales. The term $\frac{2M}{r^3}$ encompasses the classical Schwarzschild effect, dominating at large distances, and scaling directly with mass. QFs contribute through $\frac{c(3w+1)}{r^{3w+3}}$, with a state-parameter-dependent radial falloff that provides a distinctive observational signature for different models. The pure LQG correction $\frac{\alpha B^2}{2r^4}$ represents quantum spacetime discreteness effects, where $\alpha$ quantifies the strength of quantum corrections and $B$ determines their coupling to classical geometry. The mixed terms $\frac{2\alpha BM}{r^5}$ and $\frac{3\alpha M^2}{r^6}$ reveal how quantum gravity fundamentally influences the mass-curvature relationships at small scales, instead of simply contributing to classical effects. These terms create distinct regions around the BH—classical at large distances, QF-dominated at intermediate scales, and quantum-governed near the horizon—providing multiple observational windows to test theoretical predictions. The varying radial dependencies ($r^{-3}$ to $r^{-6}$) enable potential discrimination between competing theories through precision lensing measurements at different impact parameters. Most significantly, this hierarchical structure demonstrates how gravitational lensing might reveal the transition from classical to quantum gravity around LQG BHs.

To compute the deflection angle, we apply the GBTm by considering a domain bounded by the light ray trajectory and a reference curve at infinity. As $\mathcal{R} \rightarrow \infty$, Eq. (\ref{eq:gbt_main}) transforms into:
\begin{equation}\label{eq:gbt_limit}
\lim_{\mathcal{R} \to \infty}\left[\iint_{\mathcal{S}_{R}} K dS + \oint_{\partial \mathcal{S}_{\mathcal{R}}} \kappa_g ds\right] = \pi - \sum_{i=1}^{n} \epsilon_i,
\end{equation}
where we have used $\chi(\mathcal{S}_{\mathcal{R}}) = 1$ for a simply connected region.

For a domain bounded by a light ray and a circular arc at radius $\mathcal{R}$, the exterior angles sum to $\pi$ when $\mathcal{R} \to \infty$ \cite{ism06,ism07}. Thus, Eq. (\ref{eq:gbt_limit}) reduces to:
\begin{equation}\label{eq:gbt_reduced}
\iint_{\mathcal{S}_{R}} K dS + \oint_{\partial \mathcal{S}_{\mathcal{R}}} \kappa_g ds = 0.
\end{equation}

The boundary $\partial \mathcal{S}_{\mathcal{R}}$ consists of two components: the light ray trajectory $C_R$ and the circular arc $C_{\mathcal{R}}$ at radius $\mathcal{R}$. For the light ray trajectory, $\kappa_g = 0$ since it is a geodesic of the optical metric. For the circular arc, as $\mathcal{R} \to \infty$, we have $\kappa_g(C_{\mathcal{R}}) \to \frac{1}{\mathcal{R}}$ and $ds = \mathcal{R}d\phi$, which gives:
\begin{equation}\label{eq:geodesic_curvature_integral}
\int_{C_{\mathcal{R}}} \kappa_g ds = \int_{0}^{\Phi} d\phi = \Phi,
\end{equation}
where $\Phi = \pi + \hat{\alpha}_{def}$, with $\hat{\alpha}_{def}$ representing the deflection angle.

Substituting into Eq. (\ref{eq:gbt_reduced}), we obtain:
\begin{equation}\label{eq:deflection_integral_1}
\hat{\alpha}_{def} = -\iint_{\mathcal{S}_{R}} K dS - \pi = -\int_{0}^{\pi} \int_{r(\phi)}^{\infty} K \sqrt{\det g^{\text{opt}}} dr d\phi - \pi,
\end{equation}
where $r(\phi)$ describes the light ray trajectory.

Within the weak deflection limit, the light ray path can be approximated as $r(\phi) = \frac{b}{\sin\phi}$, where $b$ is the impact parameter \cite{ism08,ism09}. The area element in the optical metric is:
\begin{equation}\label{eq:area_element}
dS = \sqrt{\det g^{\text{opt}}} dr d\phi = \frac{r}{\sqrt{f(r)}} dr d\phi.
\end{equation}

Therefore, the deflection angle is given by:
\begin{equation}\label{eq:deflection_integral_2}
\hat{\alpha}_{def} = -\int_{0}^{\pi} \int_{\frac{b}{\sin\phi}}^{\infty} K \frac{r}{\sqrt{f(r)}} dr d\phi - \pi.
\end{equation}

Substituting the Gaussian curvature from Eq. (\ref{eq}) and integrating, we obtain:
\begin{equation}\label{eq}
\hat{\alpha}_{def} \approx \frac{4M}{b} + \frac{c\pi(3w+2)}{2b^{3w+1}} + \frac{\pi\alpha B^2}{4b^2} + \frac{2\pi\alpha BM}{b^3} + \frac{15\pi\alpha M^2}{4b^4} + \mathcal{O}\left(\frac{1}{b^5}\right).
\end{equation}

Our analytical expression for the deflection angle reveals a hierarchical structure of gravitational lensing effects around BHs surrounded by quintessence in LQG. The classical Schwarzschild term $\frac{4M}{b}$ establishes the fundamental light bending proportional to mass and inversely to impact parameter \cite{ism15}, representing GR's baseline prediction. This foundation is modified by the QF contribution $\frac{c\pi(3w+2)}{2b^{3w+1}}$ \cite{ism16,ism17}, whose radial dependence varies with the state parameter $w$, creating unique lensing signatures that depend on the exotic matter model. The quantum gravitational corrections introduce three progressively higher-order terms: $\frac{\pi\alpha B^2}{4b^2}$ representing pure LQG effects, $\frac{2\pi\alpha BM}{b^3}$ showing how quantum geometry interacts with mass, and $\frac{15\pi\alpha M^2}{4b^4}$ capturing higher-order mass-quantum interactions \cite{ism18,ism19}. These terms create distance-dependent regimes around BHs where different physics dominates—classical effects at large impact parameters, QF effects at intermediate scales, and LQG corrections becoming measurable only in the strong-field regime \cite{ism20}. 

\begin{figure}[ht!]
\centering
\includegraphics[width=0.8\textwidth]{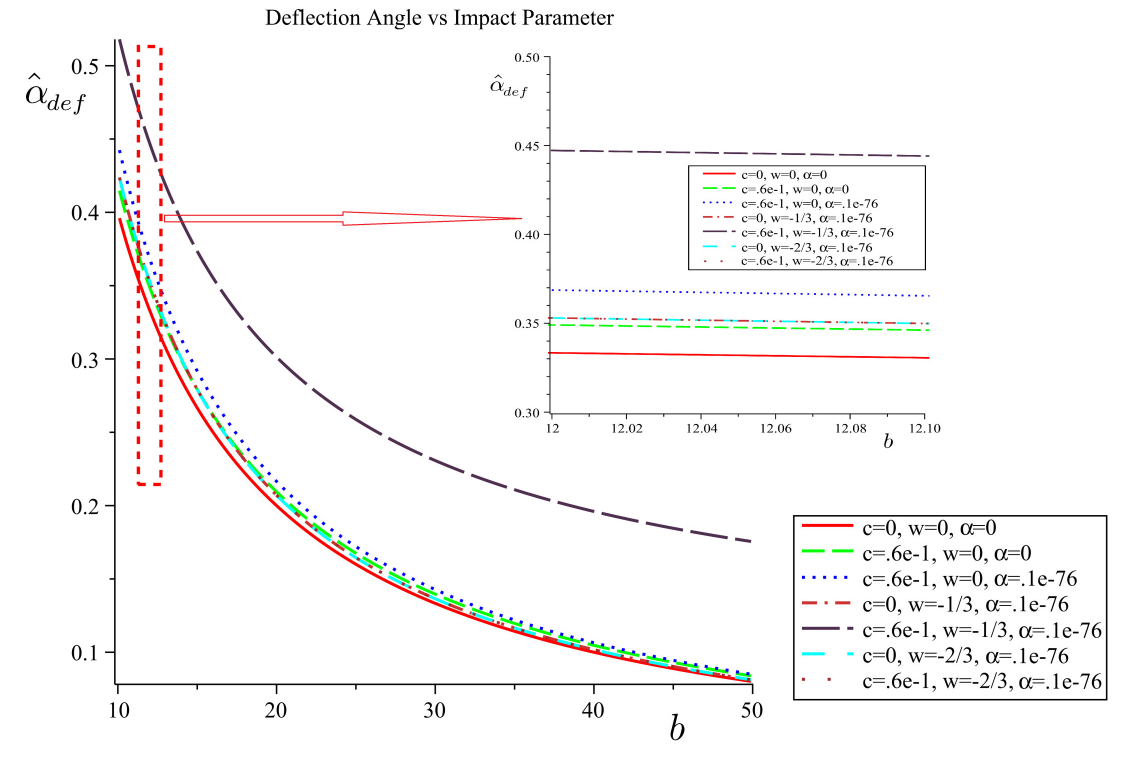}
\caption{Deflection angle $\hat{\alpha}_{def}$ as a function of impact parameter $b$ for LQG BHs surrounded by QF. The main plot shows the behavior for a wide range of impact parameters ($10 \leq b \leq 50$), while the inset provides a magnified view of the deflection angle at large impact parameters ($12 \leq b \leq 12.1$). Different curves represent various combinations of normalization factor $c$, state parameter $w$, and LQG parameter $\alpha$. The solid red line shows the classical Schwarzschild BH case ($c=0$, $w=0$, $\alpha=0$), serving as a baseline for comparison. All plots use $M=1$ and $B=1\times10^{38}$.}
\label{fig:deflection}
\end{figure}

To visualize the impact of various parameters on the gravitational lensing behavior, we present in Fig.~\ref{fig:deflection} the deflection angle $\hat{\alpha}_{def}$ as a function of impact parameter $b$ for several parameter configurations. The BH mass and LQG coupling parameter are fixed at $M=1$ and $B=1\times10^{38}$ for all curves, allowing us to isolate the effects of QF state parameter $w$, normalization factor $c$, and quantum gravity parameter $\alpha$. The curves demonstrate that while quantum corrections dominate at small impact parameters, QF effects can significantly influence deflection angles even at larger distances where classical GR would normally dominate. The figure demonstrates how the interplay between LQG corrections and QF significantly modifies light deflection around BHs across different distance scales. The baseline Schwarzschild case ($c=0$, $w=0$, $\alpha=0$) is represented by the solid red line, showing the classical $1/b$ dependence predicted by GR. When only QF effects are present ($c=6\times 10^{-1}$, $w=0$, $\alpha=0$; green dashed line), we observe a slight enhancement of the deflection angle due to the additional attractive matter distribution. The introduction of minimal LQG corrections ($\alpha=10^{-76}$) further increases the deflection angle (blue dotted line), with quantum effects becoming more pronounced at smaller impact parameters. Most notably, the combination of QF with $w=-1/3$ and LQG corrections (purple dashed line) produces the most significant deviation from the Schwarzschild case. This behavior can be understood from our analytical expression in Eq.~(\ref{eq}), where the QF term with $w=-1/3$ introduces almost a logarithmic-like dependence on $b$ rather than a power-law decay, causing its effects to persist at large distances. The magnified inset highlighting the region $12 \leq b \leq 12.1$ clearly demonstrates this persistent enhancement. For the case of quintessence with $w=-2/3$ (cyan and brown lines), the deflection enhancement is intermediate between the $w=0$ and $w=-1/3$ cases, consistent with our theoretical prediction that the effect strengthens as $w$ approaches $-1/3$. These distinct behaviors provide potentially observable signatures that could distinguish between different QF models in future high-precision lensing measurements. These numerical results confirm our analytical findings and highlight how gravitational lensing observations could potentially constrain both the properties of exotic matter surrounding BHs and quantum gravity effects. 

Recent observations by the EHT \cite{isz28} and GRAVITY instrument \cite{ism11} have begun constraining quintessence parameters, while the detection of quantum gravity terms awaits more precise future measurements, potentially offering an observational window into quantum geometry through gravitational lensing. This multi-scale structure demonstrates how precise deflection angle measurements could differentiate between competing theoretical models and potentially reveal the transition between classical and quantum gravity.

}

\section{Conclusions and Summary} \label{isec7}
{\color{black}
In this study, we conducted a comprehensive investigation of static, spherically symmetric BHs within the LQG framework surrounded by QFs. Our research addressed a significant gap in the literature by examining the interplay between quantum gravitational effects and exotic matter, revealing distinctive features that emerge from this combination.

We began our analysis by establishing the metric function for LQG BHs with QF, given by Eq.~(\ref{aa2}), incorporating both quantum corrections through parameters $\alpha$ and $B$, and quintessence effects through parameters $c$ and $w$. Our numerical analysis of the horizon structure, presented in Table~\ref{istab}, revealed several significant findings. Most notably, while classical BHs ($\alpha=0$) maintain a single horizon regardless of quintessence parameters, the introduction of quantum corrections ($\alpha=1\times 10^{-77}$) fundamentally altered this picture by consistently producing a second inner horizon. The most remarkable configuration emerged with $w=-2/3$, $c=0.06$, and $\alpha=1\times 10^{-77}$, yielding a triple-horizon structure with horizons at approximately $r=0.66$, $r=1.58$, and $r=14.43$. This represents a significant departure from both classical Schwarzschild geometry and previously studied quantum-corrected BHs. Namely, our work demonstrated how the combined effects of quantum corrections and quintessence can produce spacetime structures fundamentally different from either pure LQG BHs or classical BHs with quintessence. The triple-horizon structure, in particular, represents a novel feature emerging from this combination, with potential implications for the global causal structure of spacetime and the fate of horizons in the presence of both quantum corrections and exotic matter. Our visualization of these geometries through embedding diagrams in Fig.~\ref{fig:isfull_embedding} provided geometric intuition for these horizon structures. The diagrams clearly demonstrated the throat structure characteristic of quantum-corrected BHs, while also showing how the distant third horizon in the triple-horizon case creates an extended region where the radial coordinate behaves timelike. This behavior was further elucidated in Fig.~\ref{figa0}, showing the metric function $f(r)$ with its characteristic intersections with the horizontal axis corresponding to the horizons.

The geodesic analysis performed in Section~\ref{isec3} yielded important insights into particle dynamics around these hybrid spacetimes. We derived the effective potential for both null and timelike geodesics in Eqs.~(\ref{cc1}) and (\ref{dd1}), respectively, showing how quantum corrections and quintessence effects modify particle trajectories. For photons, we determined the photon sphere radius through Eq.~(\ref{cc5}), which lacks an analytical solution but was computed numerically for various parameter combinations. The force acting on photon particles, given by Eq.~(\ref{cc8}), demonstrated a complex dependence on both quantum and quintessence parameters, particularly evident in the specific case of $w=-2/3$ presented in Eq.~(\ref{cc9}). For timelike particles, we calculated the conserved angular momentum in Eq.~(\ref{dd3}) and particle energy in Eq.~(\ref{dd4}), revealing how both LQG and quintessence parameters influence the stability and characteristics of circular orbits. Particularly interesting was our analysis of orbital velocity in Eq.~(\ref{dd7}) and circular speed at large distances in Eq.~(\ref{dd13}), which has implications for galactic rotation curves and potentially offers an alternative explanation for observed anomalies usually attributed to dark matter \cite{izz01,izz02}.

Our investigation of BH shadows in Section~\ref{isec4} provided potentially observable consequences of these hybrid spacetimes. The equation for the photon sphere radius with $w=-2/3$, given by Eq.~(\ref{eps1}), allowed us to determine shadow properties for various parameter combinations. Fig.~\ref{figa1} illustrated how the photon radius varies with different parameters, decreasing with increasing $\alpha$ and $B$, but increasing with $c$. The shadow radii, computed using Eq.~(\ref{shadeq1}) and tabulated in Table~\ref{taba3}, showed similar dependencies. Fig.~\ref{figa2} provided comprehensive illustrations of these parameter dependencies, while Fig.~\ref{ps25} visualized the actual shadows as would be seen by distant observers. These findings suggest that shadow observations by facilities like the EHT might potentially constrain both quantum gravity and quintessence parameters simultaneously.

Section~\ref{isec5} addressed the stability of these BHs through scalar perturbation analysis. We derived the scalar perturbative potential in Eq.~(\ref{ff7}), which encodes both quantum and quintessence effects. The QNM frequencies, computed using the WKB approximation and presented in Tables \ref{taba13}–\ref{taba15}, consistently showed negative imaginary parts, confirming the stability of these hybrid BHs against scalar and electromagnetic perturbations within the examined parameter ranges. Figures \ref{realRo}–\ref{realG} illustrated how these frequencies vary with different parameters, revealing that increasing quantum parameter $\alpha$ leads to higher oscillation frequencies but slower damping rates, while increasing quintessence strength $c$ produces the opposite effect. These distinctive spectral signatures could potentially be detected in gravitational wave observations, providing another observational window into quantum gravity and exotic matter physics \cite{izz03}.

Our analysis of gravitational lensing in Section~\ref{isec6} yielded particularly interesting results. Using the GBTm approach, we derived an analytical expression for the deflection angle in Eq.~(\ref{eq}), which revealed a hierarchical structure of contributions from different physical effects. The classical Schwarzschild term $\frac{4M}{b}$ was modified by quintessence through the term $\frac{c\pi(3w+2)}{2b^{3w+1}}$, whose radial dependence varies with the state parameter. Quantum corrections introduced three progressively higher-order terms: $\frac{\pi\alpha B^2}{4b^2}$ representing pure LQG effects, $\frac{2\pi\alpha BM}{b^3}$ showing quantum-mass interactions, and $\frac{15\pi\alpha M^2}{4b^4}$ capturing higher-order quantum-mass effects. Fig.~\ref{fig:deflection} illustrated these effects across different impact parameters, showing that while quantum corrections dominate at small distances, quintessence effects can significantly influence deflection angles even at larger distances. The combination of quintessence with $w=-1/3$ and LQG corrections produced the most significant deviations from classical predictions, with effects persisting even at large impact parameters.

Looking toward future research directions, several promising avenues emerge from our findings. First, extending our analysis to rotating BHs would provide more realistic models for astrophysical objects and potentially reveal new phenomenology from the interplay of spin, quantum effects, and quintessence. Second, more detailed studies of gravitational wave emissions from binary systems involving these hybrid BHs could yield distinctive waveform signatures for detection by current and future gravitational wave observatories \cite{izz05}. Third, investigating the thermodynamic properties of these BHs, particularly their Hawking radiation, entropy, and evaporation process, would shed light on how quantum corrections and quintessence together modify BH thermodynamics. Additionally, exploring dynamical processes such as BH formation and mergers in the presence of both quantum corrections and quintessence would provide insights into the cosmic evolution of these objects. }

\small

\section*{Acknowledgments}

F.A. is grateful to the Inter University Centre for Astronomy and Astrophysics (IUCAA), Pune, India, for the opportunity to hold a visiting associateship. \.{I}.~S. extends appreciation to T\"{U}B\.{I}TAK, ANKOS, and SCOAP3 for their financial assistance. Additionally, he acknowledges the support from COST Actions CA22113, CA21106, and CA23130, which have been pivotal in enhancing networking efforts.

\section*{Data Availability Statement}

In this study, no new data was generated or analyzed.

\end{document}